\documentclass[onecolumn,12pt]{article}          
\usepackage{amsmath}
\usepackage{graphicx}
\usepackage{mathptmx}      
\usepackage{float}
\usepackage[document]{ragged2e}
\usepackage[hidelinks]{hyperref}
\usepackage{geometry}
\usepackage{hyperref}
\usepackage{tikz}
\usetikzlibrary{shapes,backgrounds}
\usepackage{placeins}
\usepackage{epstopdf}
\usepackage{wasysym}
\usepackage{gensymb}
\usepackage{textcomp}
\usepackage{multirow}
\usepackage{fullpage}
\usepackage{subfig}
\usepackage{natbib}
\usepackage{xcolor}
\usepackage{subfig}

\newcommand{\triup}[1]{\tikz[baseline=0pt]{\draw[draw,thick,fill,#1](0.0,0.0) -- (0.2,0.0) -- (0.1,0.2)-- cycle;}}
\newcommand{\tridown}[1]{\tikz[baseline=0pt]{\draw[draw,thick,fill,#1](0.0,0.2) -- (0.2,0.2) -- (0.1,0.0)-- cycle;}}
\newcommand{\tstar}[5]{
\pgfmathsetmacro{\starangle}{360/#3}
\draw[#5] (#4:#1)
\foreach \x in {1,...,#3}
{ -- (#4+\x*\starangle-\starangle/2:#2) -- (#4+\x*\starangle:#1)
}
-- cycle;
}
\newcommand{\rect}[1]{\tikz[baseline=0pt]{\draw[draw,thick,fill,#1](0.0,0.0) -- (0.0,0.2) -- (0.2,0.2)-- (0.2,0.0)-- cycle;}}
\newcommand{\swrect}[1]{\tikz[baseline=0pt]{\draw[draw,thick,fill,#1](0.1,0.0) -- (0.0,0.1) -- (0.1,0.2) -- (0.2,0.1) --cycle;}}

\definecolor{green}{rgb}{0.0, 0.5, 0.0}
\definecolor{yellow}{rgb}{0.75, 0.75, 0.0}
\definecolor{magenta}{rgb}{0.75, 0, 0.75}
\definecolor{cyan}{rgb}{0.0, 0.75, 0.75}
\definecolor{pinot}{rgb}{0.64, 0.0784, 0.1765}
\definecolor{royal}{rgb}{0.0784, 0.1686, 0.549}
\definecolor{dor}{rgb}{0.773, 0.353, 0.07}

\usepackage{xcolor}
\hypersetup{
    colorlinks,
    linkcolor={red!50!black},
    citecolor={blue!50!black},
    urlcolor={blue!80!black}
}

\usepackage{multirow}
\usepackage[affil-it]{authblk}
\usepackage[english]{babel}
\usepackage[none]{hyphenat}
\usepackage{fancyhdr}
\title{An aeroacoustic investigation into the effect of self-oscillating trailing edge flaplets}
\renewcommand{\thefootnote}{\fnsymbol{footnote}}
\author[1]{Edward Talboys \thanks{Corresponding author: edward.talboys.1@city.ac.uk}}
\author[2]{Thomas F. Geyer}
\author[1]{Christoph Br\"{u}cker}
\affil[1]{City, University of London\\ Northampton Square, London, EC1V 0HB, United Kingdom}
\affil[2]{Brandenburg University of Technology Cottbus - Senftenberg\\ 03046 Cottbus, Germany}
\date{}

\begin{document}
\maketitle
\renewcommand{\thefootnote}{\arabic{footnote}}
\thispagestyle{fancy}
\justify
\begin{abstract}
The aeroacoustics of a NACA 0012 aerofoil with an array of self-oscillating flexible flaplets attached on the trailing edge has been investigated at low to moderate chord based Reynolds number (50,000 -- 350,000) and at geometric angles of attack from $\alpha_g = 0^\circ$ -- $20^\circ$. When the aerofoil is untripped, tonal peaks are observed on the baseline aerofoil. When the passive flaplets are attached to the pressure side of the aerofoil, the tonal peak is removed. If the flaplets are then placed on the suction side, the tonal peak is reduced, but not removed. It is therefore hypothesised that the flaplets on the pressure side modifies the laminar separation bubble situated on the pressure side of the aerofoil, a key mechanism for tonal noise. Throughout all cases, both tripped and untripped, a low frequency (0.1~kHz -- 0.6~kHz) noise reduction and a slight increase at higher frequencies ($>$2~kHz) is seen. This gives an average overall sound pressure level (OSPL) reduction of 1.5 -- 2~dB for the flaplets affixed to the pressure side. The cases where the tonal noise component is removed an OSPL reduction of up to 20~dB can be seen. 
\end{abstract}

\section{Introduction}
Aerofoil self-noise noise reduction is a topic which is attracting increasing interest due to the growing need and desire for `quieter' aerofoils for various engineering applications. 
The main source of this self-noise is the boundary layer -- trailing edge interaction. 
Therefore various strategies have been proposed by engineers in recent years to mitigate this. 

When the aerofoil is subjected to a laminar boundary layer, for moderate Reynolds numbers, an annoying tonal noise is present and as such a significant amount of research has been carried out in order to try to understand this phenomenon. 
The first study into tonal noise was by \citet{Paterson1972}, where they made the observation that the main tonal peak can be scaled with the freestream velocity, $U_\infty$, by a factor of 0.8 over a short range prior jumping to a higher frequency, which is commonly referred to as `laddering'. 
Once all of the `laddering' events are averaged out over a large frequency range, a scaling relationship of $U_\infty^{1.5}$, is observed and hence an empirical scaling model was proposed.
\citet{Tam1974} built off these results and proposed an aeroacoustic self-excited feedback loop which is formed by instabilities in the boundary layer on the pressure side of the aerofoil and a noise source situated in the wake. 
\citet{ARBEY1983} then expanded on Tam's feedback model by showing that Tollmein-Schlichting (T-S) waves defracting at the trailing edge, creating acoustic waves, initiate the feedback loop. 
This result initiated much more detailed investigations into the flow field around the aerofoil. 
\citet{Lowson} and \citet{McAlpine1999} showed that the tonal noise is governed by the presence of a laminar separation bubble on the pressure side of the aerofoil and that the frequency of the tonal noise is the most amplified frequency in the boundary layer by using linear stability theory. 
\citet{Desquesnes2007} carried out the first direct numerical simulation (DNS) on the tonal noise issue and they found that there is a secondary feedback loop which comes from the instabilities on the suction side of the aerofoil. 
This feedback was then thought to modulated the discrete frequencies which are evenly spaced around the main tonal peak. 
\citet{Probsting2015} then used simultaneous particle image velocimitry (PIV) and acoustic measurements to show, at very low Reynolds numbers (Re$_\text{c}$ = 30,000) the tonal noise generation is controlled by the suction side and then at higher Reynolds numbers (Re$_\text{c}$ = 230,000) the tonal noise is dominated by the pressure side. 
They also show that by tripping either side of the aerofoil, separately, the dual side feedback loop \citep{Desquesnes2007} is not a necessity in order to observe the tonal noise.  

To mitigate this aeroacoustic phenomenon many possible approaches exist, most of which are bio-inspired.
\citet{Geyer2010} investigated a wide range of aerofoils with different porosities, where the inspiration came from the `soft downy feathers' of the owl, which are well known for `silent' flight. 
They observed that if the aerofoil has any porosity, an aeroacoustic benefit was seen in the low to mid frequency range, upto 10~dB but this was very dependant on the porosity. 
However in the high frequency range there was an increase when compared to a non-porous aerofoil and this was thought to be noise originating from the surface roughness of the aerofoil. 
There was also an aerodynamic penalty, where both lift and drag forces were negatively effected compared to the non porous aerofoil. 

Another owl bio-inspired technique, trailing edge brushes, mimicking the trailing edge fringes observed on the owls feather, was investigated by \citet{Herr2007} and \citet{Finez2010}. 
A broadband noise, turbulent boundary layer -- trailing edge interaction, reduction ranging from 2--10~dB was observed by \citep{Herr2007} in the frequency range of 1--16~kHz. 
\citet{Finez2010}, then built upon this result by showing that the spanwise coherence of the shed vorticies is reduced by 25\% in the presence of trailing edge brushes. 
And have proposed that the design criteria for future brush designs, should be a ratio of coherence length to brush fibre diameter and have suggested that a ratio of 2 is optimal.
 
Another common technique that is used is trailing edge serrations.
This has been extensively researched and have shown positive aeroacoustic results in both the laminar boundary layer case \citep{Chong2010} and turbulent boundary layer case \citep{Leon2017}. 

In the present study, a flexible trailing edge consisting of an array of small elastic flaplets, mimicking the tips of bird feathers is used. 
This type of trailing edge has had very few investigations thus far. \citet{Schlanderer2013} carried out a DNS on a flat plate with an elastic compliant trailing edge and saw an aeroacoustic benefit at low and medium frequencies with an increased noise level at the Eigen frequency of the material. 
\citet{Das2015} then carried out an experimental investigation using a similar arrangement to \citet{Schlanderer2013}. 
An overall average reduction of 4~dB was seen. 
\citet{Kamps2017} then applied silicone flaplets to the trailing edge of a NACA 0010 aerofoil, and showed a reduction in tonal noise but no reduction in broadband noise.

This type of trailing edge modification has also been seen to have an aerodynamic advantages. \citet{Talboys2018} showed by using time resolved PIV on a NACA0012, that an array of oscillating flexible flaplets stabilise the shear layer at moderate Reynolds number and angle of attack. This was seen to subsequently reduce the boundary layer on the suction side of the aerofoil and was hypothesised to then lead to a reduction in drag. 
\citet{Jodin2018} used an active solid trailing edge which oscillated in a similar manner and frequency to the flaplets in the present study and \citet{Talboys2018}. 
Their investigation was focused on the wake structure and it was observed that the wake could be reduced in thickness by as much as 10\%.
An increase in lift of 2\% was also observed. 

The present study builds off the initial aeroacoustic investigation of \citet{Kamps2017} in order to provide a more in depth aeroacoustic analysis on the benefits of using a passive array of self-oscillating flexible flaplets when subjected to a laminar boundary layer and a turbulent boundary layer. The preliminary results of present work was presented at the `IUTAM Symposium on Critical flow dynamics involving moving/deformable structures with design applications, 2018' \citep{Talboys2018a}.

\section{Experimental Arrangement}
\label{sec: Exp Arr Acc}
\begin{figure}[tb]
\begin{minipage}[b]{.48 \linewidth}
\centering 
\subfloat[Schematic display of the measurement setup (top view, \texttimes ~marks the location of the single microphone)]{\label{fig: Exp Arr. Acc}
\small
\begin{tikzpicture}[scale=3.5,>=latex]
		\draw[thick,royal,fill,fill opacity=0.2] (-0.48,-0.14) rectangle (-0.28,0.14);
		\draw[royal](-0.3,-0.14) -- +(-45:0.3) node [below] {aerofoil at $z=0$~m};
		\begin{scope}[scale=0.2,xshift=-2.65cm]
			\draw[thick](-2,2) node [above right,text width=2cm] {nozzle} .. controls (-1,2) and (-0.7,0.5) .. (0,0.5);
			\draw[thick](-2,-2) .. controls (-1,-2) and (-0.7,-0.5) .. (0,-0.5);
			\shadedraw[dotted,left color=gray!40,opacity=0.8](5,0) -- (0,0.5) -- (5,1.5);
			\shadedraw[dotted,left color=gray!40,opacity=0.8](5,0) -- (0,-0.5) -- (5,-1.5);
			\draw[->](-1,0.25)--(0,0.25);
			\draw[->](-1,0)--(0,0) node [at start,left]{};
			\draw[->](-1,-0.25)--(0,-0.25);
			\draw(2.15,0) -- +(110:3.5) node [above] {core jet};
			\draw(2.2,0.7) -- +(85:3) node [above] {mixing zone};
		\end{scope}
		\draw[pinot,only marks,mark=*,mark options={scale=0.15}] plot coordinates {
		(-0.145527,0.6335)
		(-0.066774,0.236764)
		(-0.376879,0.529587)
		(-0.152297,0.193188)
		(-0.550855,0.345049)
		(-0.214634,0.120201)
		(-0.640968,0.10798)
		(-0.244295,0.028914)
		(-0.6335,-0.145527)
		(-0.236764,-0.066774)
		(-0.529587,-0.376879)
		(-0.193188,-0.152297)
		(-0.345049,-0.550855)
		(-0.120201,-0.214634)
		(-0.10798,-0.640968)
		(-0.028914,-0.244295)
		(0.145527,-0.6335)
		(0.066774,-0.236764)
		(0.376879,-0.529587)
		(0.152297,-0.193188)
		(0.550855,-0.345049)
		(0.214634,-0.120201)
		(0.640968,-0.10798)
		(0.244295,-0.028914)
		(0.6335,0.145527)
		(0.236764,0.066774)
		(0.529587,0.376879)
		(0.193188,0.152297)
		(0.345049,0.550855)
		(0.120201,0.214634)
		(0.10798,0.640968)
		(0.028914,0.244295)
		(-0.029598,0.463341)
		(-0.02879,0.088281)
		(-0.348561,0.306703)
		(-0.082782,0.042067)
		(-0.463341,-0.029598)
		(-0.088281,-0.02879)
		(-0.306703,-0.348561)
		(-0.042067,-0.082782)
		(0.029598,-0.463341)
		(0.02879,-0.088281)
		(0.348561,-0.306703)
		(0.082782,-0.042067)
		(0.463341,0.029598)
		(0.088281,0.02879)
		(0.306703,0.348561)
		(0.042067,0.082782)
		(-0.138631,0.241627)
		(-0.268883,0.072829)
		(-0.241627,-0.138631)
		(-0.072829,-0.268883)
		(0.138631,-0.241627)
		(0.268883,-0.072829)
		(0.241627,0.138631)
		(0.072829,0.268883)
		};
		\draw[black,only marks,mark=x,mark options={scale=0.4},opacity=0.1] plot (-0.244295, 0.028914);
		\draw[pinot](0.45,0.5) node [above right,text width=2.5cm] {microphones at $z=0.71$~m};
		\draw[->] (-0.932,-0.72) -- (-0.932,0.8) node [right] {$y$ in m};
		\draw[->] (-0.932,-0.72) -- (0.9,-0.72) node [below] {$x$ in m};
		\foreach \x in {-0.53,-0.28,0,0.5}
			 \draw (\x,-0.7) -- +(0,-0.04) node [below] {\x};
		\foreach \y in {-0.2,-0.1,0,0.1,0.2} 
			 \draw (-0.912,\y) -- +(-0.04,0) node [left] {\y};		
	\end{tikzpicture}}\normalsize
\end{minipage}
\begin{minipage}[b]{.48 \linewidth}	\label{fig: Aerofoil Pic}
\centering
\subfloat[Photo of the NACA 0012 aerofoil with the flaplets adhered on the trailing edge.]{\includegraphics[scale=0.6]{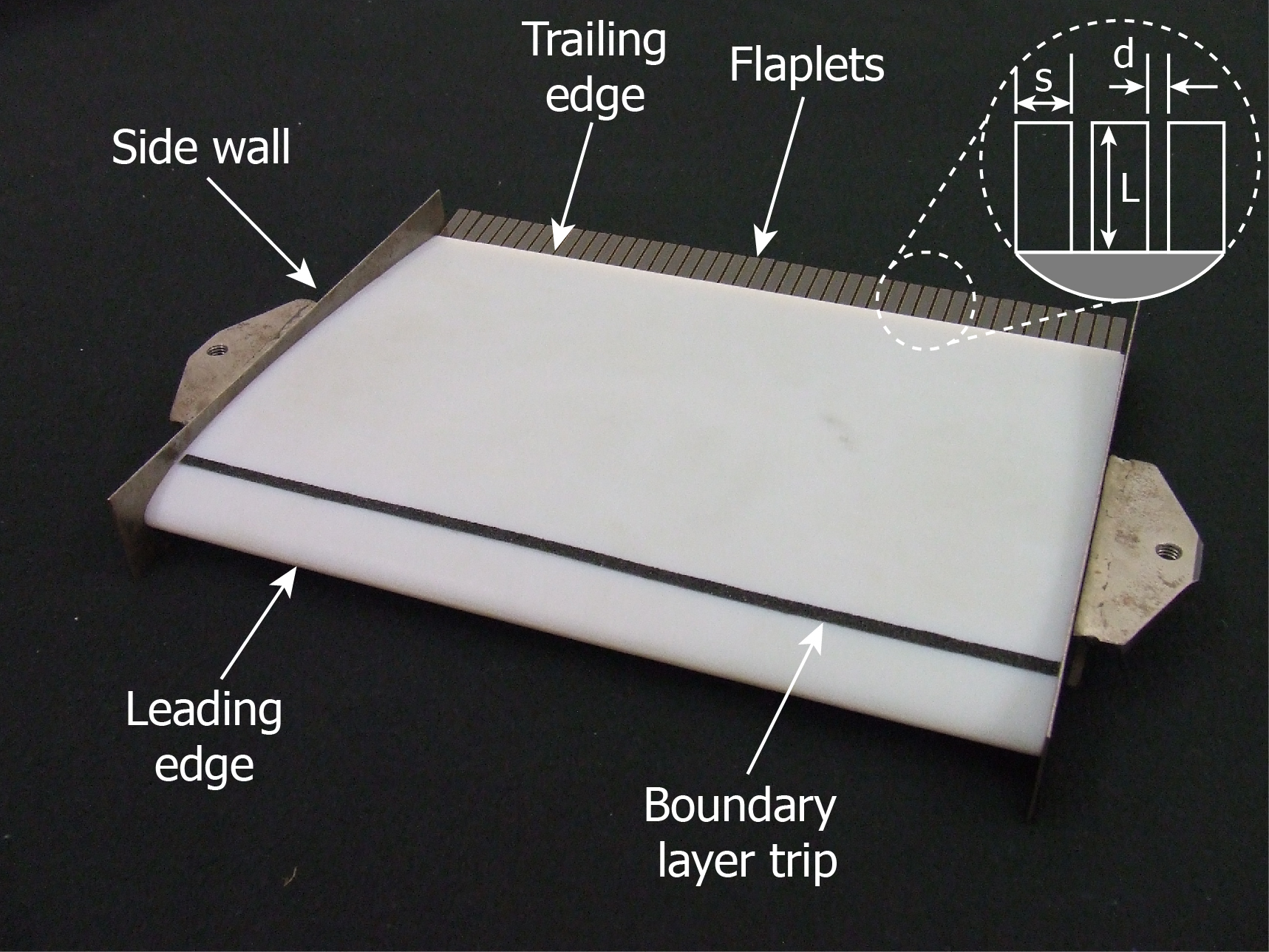}}
\end{minipage}\medskip\par
\begin{minipage}{\linewidth} \label{fig: Aerofoil Flap}
\centering
\subfloat[Showing the two different flaplet placements at an angle of incidence, $\alpha$. Pressure side placement (top) and suction side placement (bottom)]{\includegraphics[scale=.9]{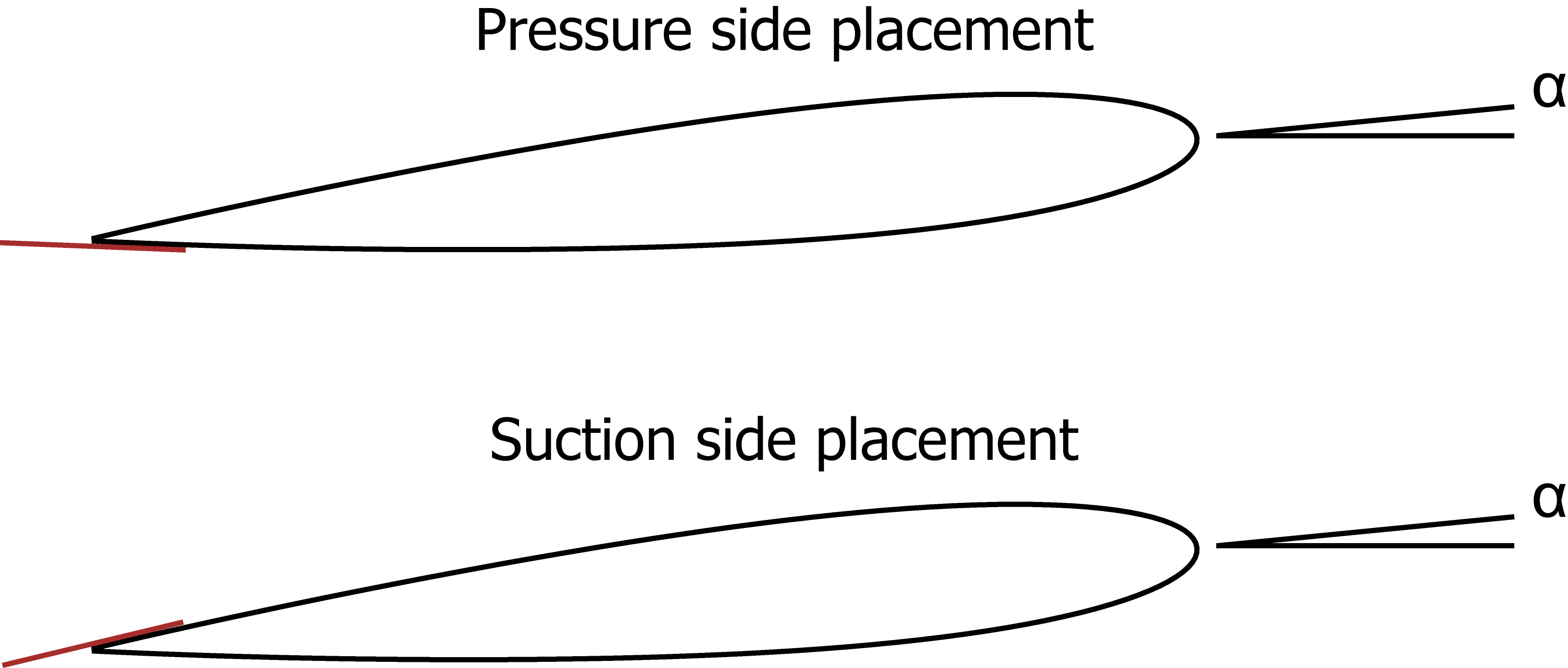}}
\end{minipage}
\caption{Experimental set-up}
	\label{fig: Set-up}
\end{figure}
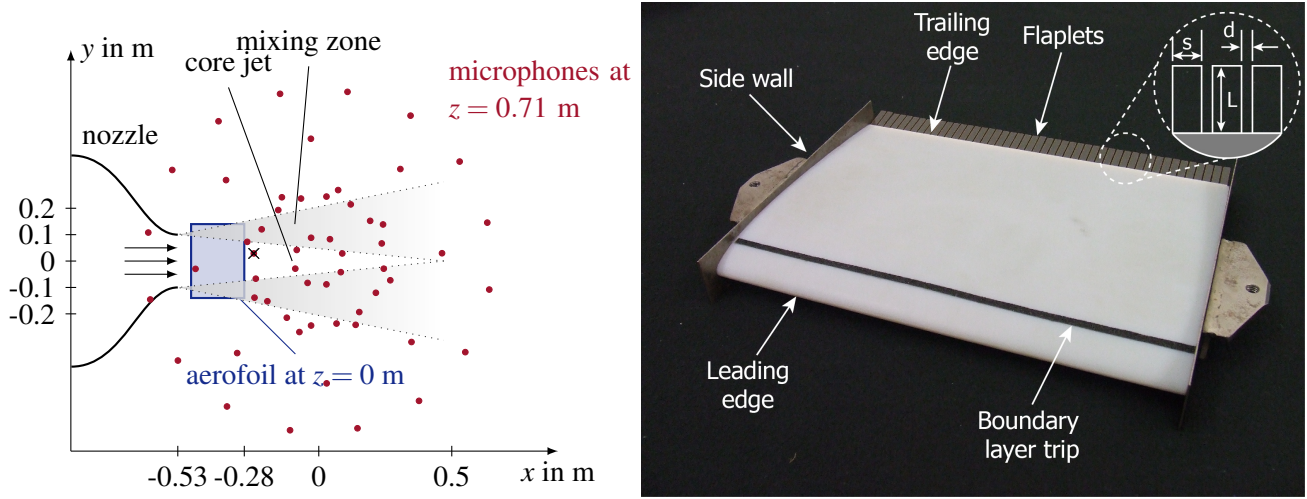
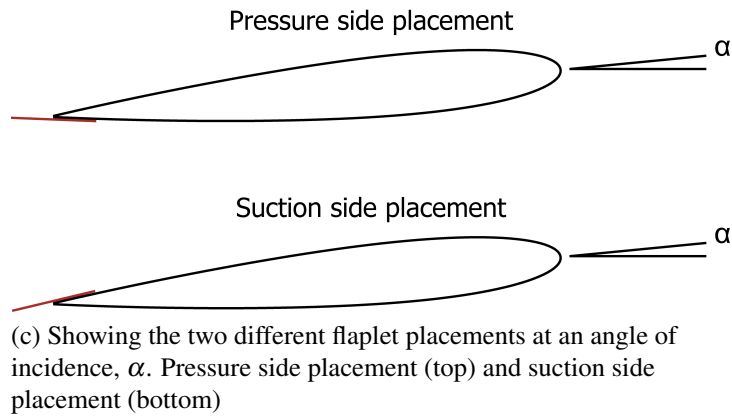

The aerofoil used for the present study was a NACA 0012, with a chord ($c$) of 0.2~m and a span ($b$) of 0.28~m. 
The model was 3D printed in two section, where the adjoining plane was the chord line of the aerofoil to allow for equal surface finish on both sides of the aerofoil. 
Manufacturing the aerofoil using this technique allows the trailing edge to be minimal, to reduce trailing edge bluntness, and in this instance the thickness was 0.5~mm with a solid angle of 16$^\circ$. 
The flexible trailing edge flaplets were manufactured, using a laser cutter, from a thin polyester film (see table \ref{table: Flaplet Dimensions} for dimensions and material properties). 
The flaplets were attached to the aerofoil using a thin strip of double sided tape, and were placed such that the free ends were orientated downstream at 1.1c and are allowed to freely oscillate at their Eigen frequency in the flow field. 
The Eigen frequency was determined to be 107~Hz in a previous study \citep{Talboys2018}, by using cantilever beam theory.

In order to investigate the flaplets effect in the presence of both a turbulent and laminar boundary layer, a 0.15~mm thick boundary layer trip was applied at 0.2c on both the suction and pressure sides of the aerofoil. 
The flaplets were also placed on both sides of the aerofoil, separately, in order to see if the there is any effect depending on whether the flaplets were placed on the pressure or suction side. 
 
\begin{table}
\centering
\begin{tabular}{c|c|c|c|c|c|c}
\multirow{2}{*}{Length (L)} & \multirow{2}{*}{Width (s)} & \multirow{2}{*}{\begin{tabular}[c|]{@{}c@{}}Inter\\ spacing (d)\end{tabular}} & \multirow{2}{*}{Thickness} & \multirow{2}{*}{Density} & \multirow{2}{*}{\begin{tabular}[c|]{@{}c@{}}Young's\\ Modulus\end{tabular}} & \multirow{2}{*}{\begin{tabular}[c|]{@{}c@{}}Eigen\\ frequency\end{tabular}} \\
&	&	&	&	&	&	\\ \hline
20~mm	& 5~mm	& 1~mm	& 180~$\mu$m	& 1440~kg/m$^3$	& 3.12~GPa	& 107~Hz
\end{tabular}
\caption{Flaplet Dimension and Material Properties}\label{table: Flaplet Dimensions} 
\end{table}     

The acoustic measurements took place in the small aeroacoustic open jet wind tunnel \citep{Sarradj2009} at the Brandenburg University of Technology in Cottbus, with a setup similar to that used in \citet{Geyer2010}. The wind tunnel was equipped with a circular nozzle with a contraction ratio of 16 and an exit diameter of 0.2~m. With this nozzle, the maximum flow speed is in the order of 90~m/s and at 50~m/s, the turbulence intensity in front of the nozzle is below 0.1~\%. For the present study the chord based Reynolds number was varied from 50,000 -- 350,000 and the geometric angle of attack, $\alpha_g$, was varied from $\alpha_g$ = 0$^\circ$ -- 20$^\circ$. As the wind tunnel is open jet, a correction factor is commonly applied to the angle of attack. This correction factor was introduced by \citet{Brooks1986} who used lifting line theory to account for the deflection induced by the open jet boundary conditions. 
However due to the small jet width to aerofoil chord ratio ($b/c = 1.4$), the correction factor should be used with caution \citep{Moreau2003} and as such has not been used to indicate the angle in the present study.
All angles, unless otherwise stated, are therefore the geometric angles of attack ($\alpha_g$).
During measurements, the wind tunnel test section is surrounded by a chamber with absorbing walls on three sides, which lead to a quasi anechoic environment for frequencies above 125~Hz.

For the measurements, the aerofoil is positioned at a distance of 0.05~m downstream from the nozzle. The tips of the aerofoil are attached to a six component wind tunnel balance to simultaneously measure the integral aerodynamic forces. Since the span of the aerofoil ($b$~=~0.28~m) exceeds the nozzle diameter, no aerodynamic noise is generated at the tips or the lateral mountings. A schematic of the setup is shown in Figure~\ref{fig: Set-up}.

The acoustic measurements were performed using a planar microphone array, consisting of 56 1/4th inch microphone capsules flush mounted into an aluminium plate with dimensions of 1.5~m~\texttimes~1.5~m (see \citet{Sarradj2010}). The microphone layout is included in Figure~\ref{fig: Set-up}. The aperture of the array is 1.3~m. The array was positioned out of the flow, in a distance of 0.71~m above the aerofoil.

Data from the 56 microphones were recorded with a sampling frequency of 51.2~kHz and a duration of 60~s using a National Instruments 24~Bit multichannel measurement system. To account for the refraction of sound at the wind tunnel shear layer, a correction method was applied that is based on ray tracing \citet{Sarradj2017}. 
In post processing, the time signals were transferred to the frequency domain using a Fast Fourier Transformation, which was done blockwise on Hanning-windowed blocks with a size of 16384 samples and 50~\% overlap. 
This lead to a small frequency spacing of only 3.125~Hz. 
The resulting microphone auto spectra and cross spectra were averaged to yield the cross spectral matrix. 
This matrix was further processed using the CLEAN-SC deconvolution beamforming algorithm \citet{Sijtsma2007}, which was applied to a two-dimensional focus grid parallel to the array and aligned with the aerofoil. 
The grid has a streamwise extent of 0.5~m, a spanwise extent of 0.4~m and an increment of 0.005~m. The outcome of the beamforming algorithm is a two-dimensional map of noise source contributions from each grid point, a so-called sound map. 
In order to obtain spectra of the noise generated by the interaction of the turbulent boundary layer with the trailing edge of the aerofoil, a sector was defined that only contains the noise source of interest. 
The chosen sector has a chordwise extent of 0.2~m and a spanwise extent of 0.1~m. 
Thus, spectra of the noise generated by this mechanism are derived by integrating all noise contributions from within this sector, while all potential background noise sources (such as the wind tunnel nozzle or the aerofoil leading edge) are excluded from the integration. 
The resulting sound pressures were then converted to sound pressure levels $L_p$ $re$ 20~$\mu$Pa and 6~dB were subtracted to account for the reflection at the rigid microphone array plate.

In addition to the beamforming results, auto spectra of a single array microphone close to the aerofoil trailing edge were analysed. 
The microphone position is highlighted in fig.\ref{fig: Exp Arr. Acc}.

\section{Results}
\subsection{Theoretical Comparison}
Brooks, Pope and Marcolini \citep{Brooks1989} created a semi-empirical model that aims to predict the aerofoil self-generated noise by breaking it down into five main components; laminar boundary layer -- trailing edge interaction (LBL--TE), turbulent boundary layer -- trailing edge interaction (TBL--TE, both on suction and pressure sides), separated flow noise, trailing edge bluntness and tip vortex noise. As the aerofoil used in the present study is bounded by two end plates, the tip vortex noise is not considered. In order to use this model to predict and analyse the noise sources the open source software, NAFNoise \citep{Moriarty2005}, was used. NAFNoise uses a panel method, Xfoil, to calculate the necessary boundary layer parameters for the model and has an additional feature which uses a simplified version of the Guidati model, to calculate the additional noise induced from a turbulent inlet flow. As mention in section \ref{sec: Exp Arr Acc}, the inflow turbulence is low for the present experimental set-up; nonetheless this has still been accounted for in the prediction. 

\begin{figure}[h!] 	
\begin{minipage}{.48\linewidth}
\centering
\subfloat[BPM Empirical Model]{\label{fig: BPM-a}\includegraphics[scale=.6]{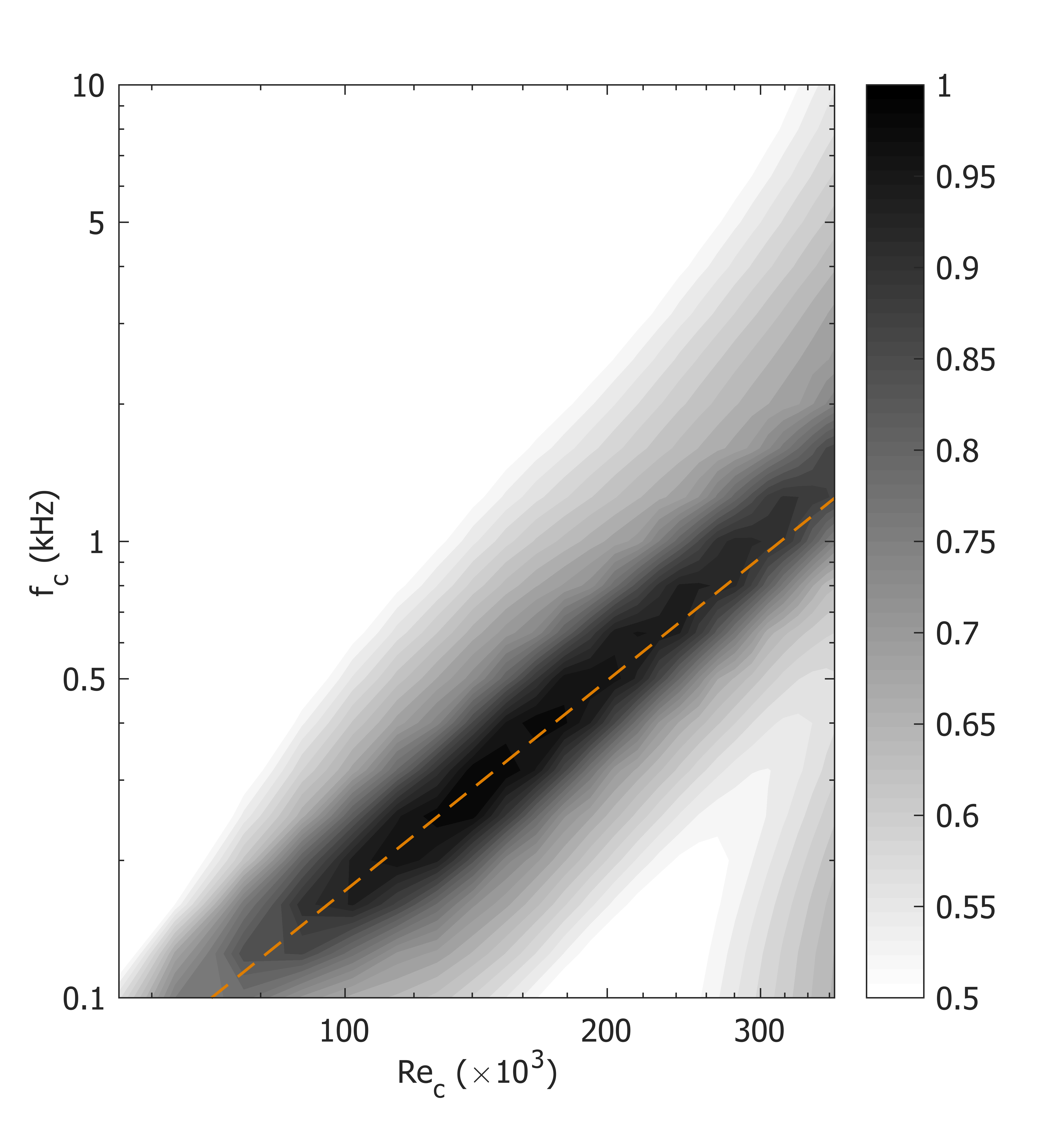}}
\end{minipage}
\begin{minipage}{.48\linewidth}
\centering
\subfloat[Single microphone experimental result]{\label{fig: BPM-b}\includegraphics[scale=.6]{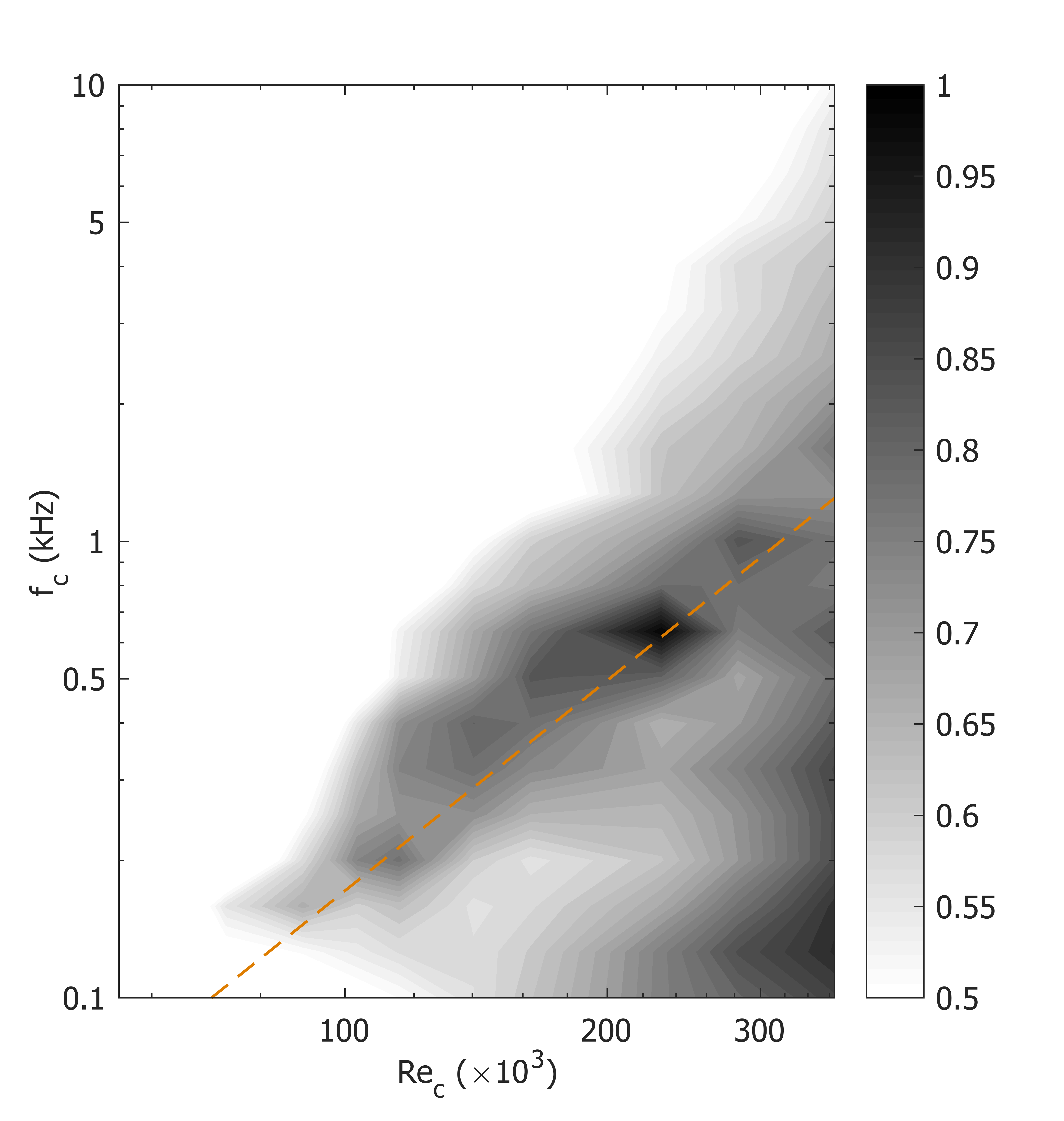}}
\end{minipage}\par
\caption{Contours of normalised third octave band SPL across the Reynolds number range studied at $\alpha_g = 0^\circ$. The contours are normalised by their respective maximum SPL. The \protect\begin{tikzpicture}
\protect\tikz[baseline=-2pt] \protect\draw[solid,very thick,color=dor] (0.0,0.0) -- (0.15,0.0) (0.2,0.0) -- (0.35,0.0) (0.4,0.0) -- (0.55,0.0);
\protect\end{tikzpicture} shows the $\sim Re_c^{1.5}$ trend line observed by \citet{Paterson1972}}
\label{fig: BPM Contours}
\end{figure}

Figure \ref{fig: BPM Contours} shows the comparison of the BPM model with the baseline experimental results at $\alpha_g = 0^\circ$, for all of the Reynolds numbers tested. The contours have been normalised by their respective maximum SPL in order to compare the overall trends. Immediately it can be seen that in fig. \ref{fig: BPM-a}, there is a clear trend of increasing tonal peak with Reynolds number. This trend corresponds well with the empirical Re$^{1.5}$ scaling from \citet{Paterson1972}. As the results from the BPM model are in third octave bands, the `laddering' effect, which scales as Re$_\text{c}^{0.8}$, cannot be seen and only the average effect is observed. In the experimental results, fig. \ref{fig: BPM-b}, the trend is also clearly visible and hence the BPM model can be used for the current experimental set-up over a wide range of Reynolds numbers to help to understand the noise sources.   

\begin{figure}[h!] 	
\begin{minipage}{.48\linewidth}
\centering
\subfloat[$\alpha_g = 0^\circ$]{\label{fig: BPM-0deg}\includegraphics[scale=.6]{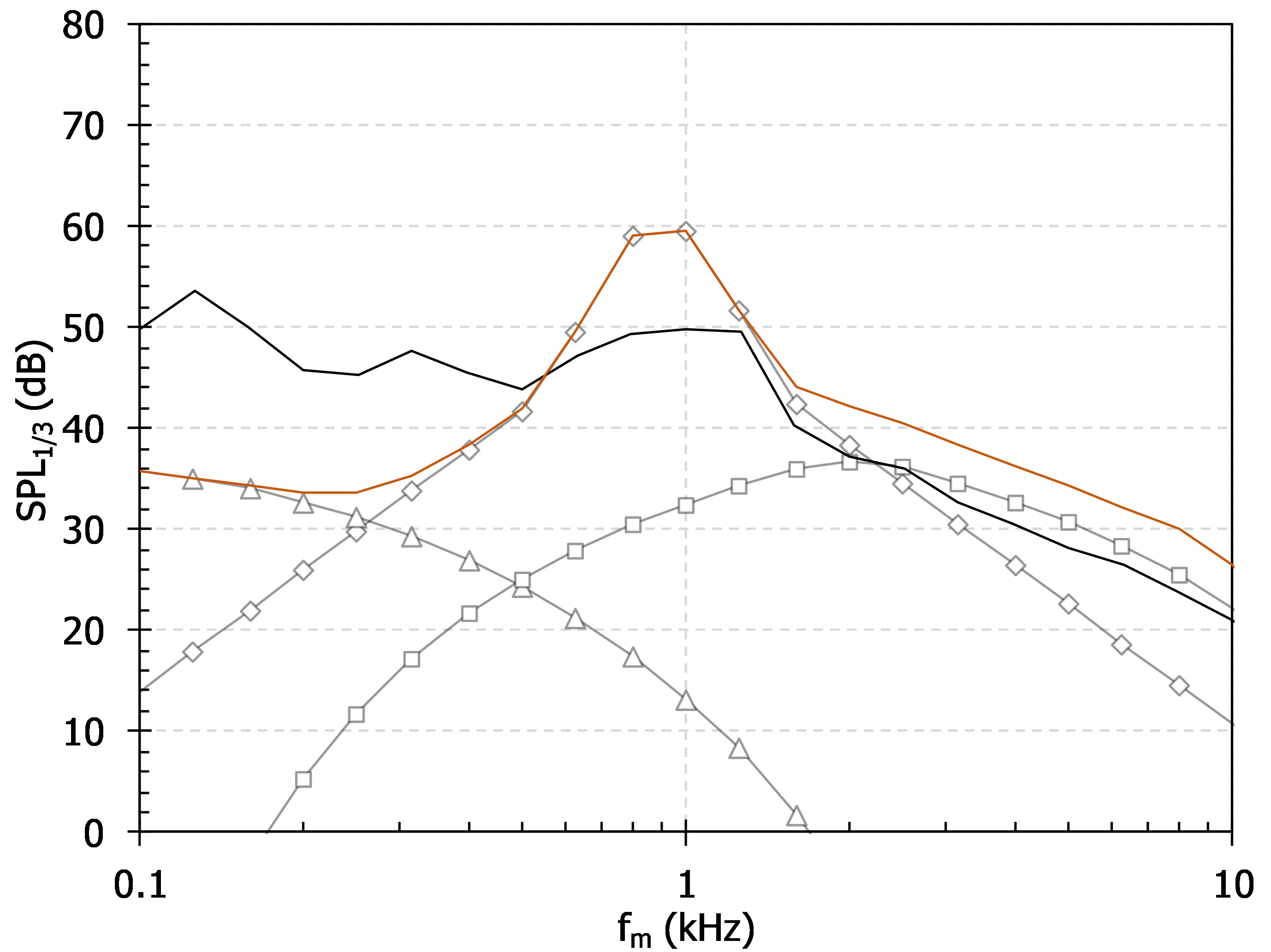}}
\end{minipage}
\begin{minipage}{.48 \linewidth}
\centering
\subfloat[$\alpha_g = 10^\circ$]{\label{fig: BPM-10deg}\includegraphics[scale=.6]{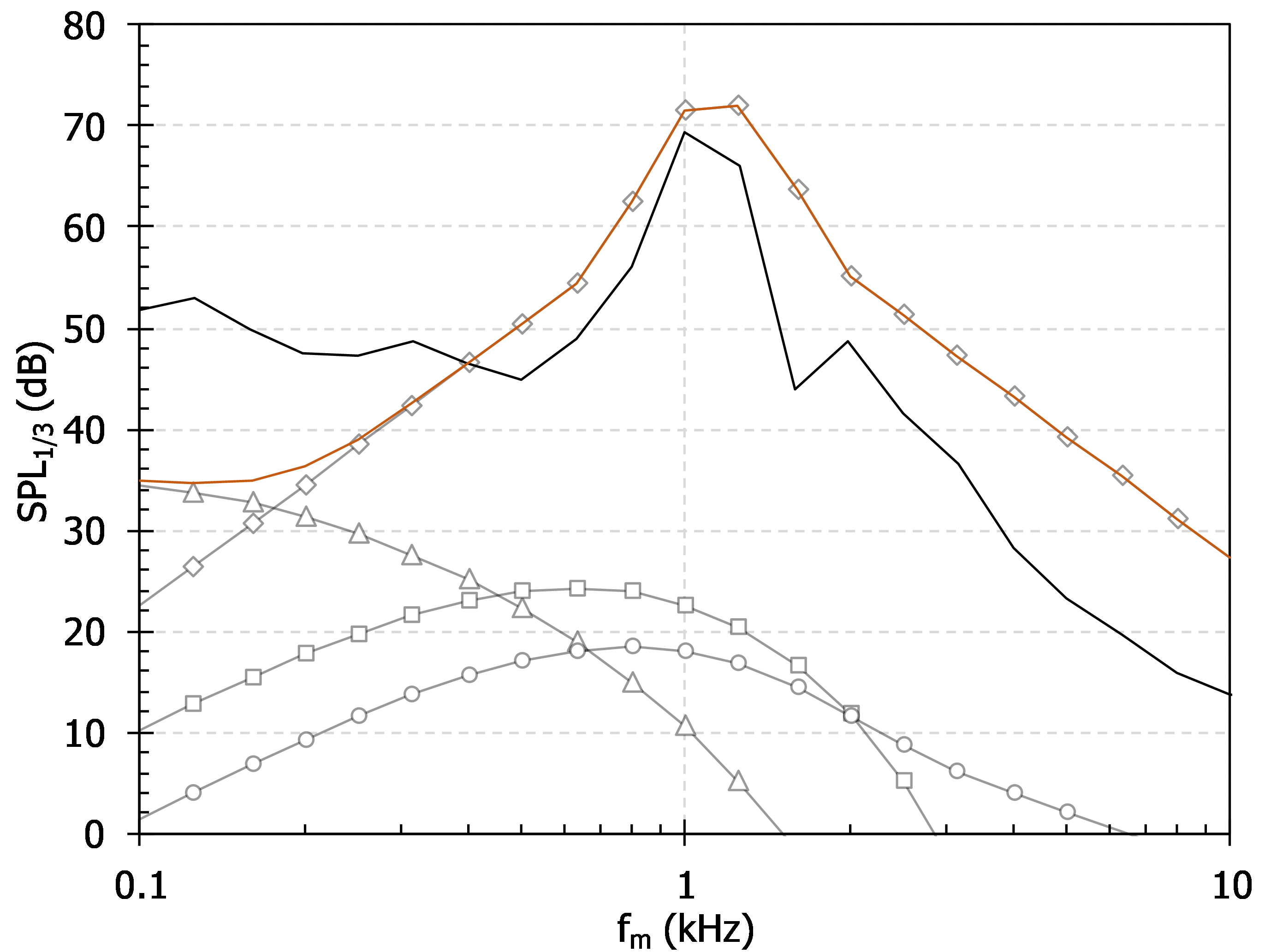}}
\end{minipage}\par
\caption{Comparison of the third octave sound pressure level (SPL$_{\protect\text{1/3}}$) at $Re_c$ = 300,000 between experimental result and the BPM prediction model. Where 
\protect\begin{tikzpicture}
\protect\tikz[baseline=-2pt] \protect\draw[solid,very thick,color=black] (0.0,0.0) -- (0.3,0.0);
\protect\end{tikzpicture} is the experimental case and 
\protect\begin{tikzpicture}
\protect\tikz[baseline=-2pt] \protect\draw[solid,very thick,color=dor] (0.0,0.0) -- (0.3,0.0);
\protect\end{tikzpicture} is the BPM model, which is a summation of:
\protect\begin{tikzpicture}
\protect\swrect{color=gray,fill=white};
\protect\end{tikzpicture} LBL,
\protect\begin{tikzpicture}
\protect\rect{color=gray,fill=white};
\protect\end{tikzpicture} pressure side TBL,
\protect\begin{tikzpicture}
\protect\tikz[baseline=0pt] \protect\draw[thick,color=gray,fill=white] (0.1,0.1) circle (0.1);
\protect\end{tikzpicture} suction side TBL and
\protect\begin{tikzpicture}
\protect\triup{color=gray,fill=white};
\protect\end{tikzpicture} inflow noise.}
\label{fig: BPM Comparison}
\end{figure}

Figure \ref{fig: BPM Comparison}, shows the BPM prediction against the experimental results for the baseline case and both of the flaplet orientation cases for one Reynods number, 300,000 and at two different geometric angles of attack. In general the BPM model can predict the frequency of the tonal peak at both angles of attack, however the magnitude is over predicted. The LBL-TE noise is the dominating source in both cases, which is to be expected as the transition is not forced. In the 0$^\circ$ case (fig. \ref{fig: BPM-0deg}), it can be seen that in the higher frequency range the pressure/suction side TBL (only the pressure side TBL has been plotted in fig. \ref{fig: BPM-0deg}) becomes the dominating noise source, but the influence is on the overall noise level is small in relation to the LBL-TE tonal peak.  

\subsection{Single Microphone Measurements}
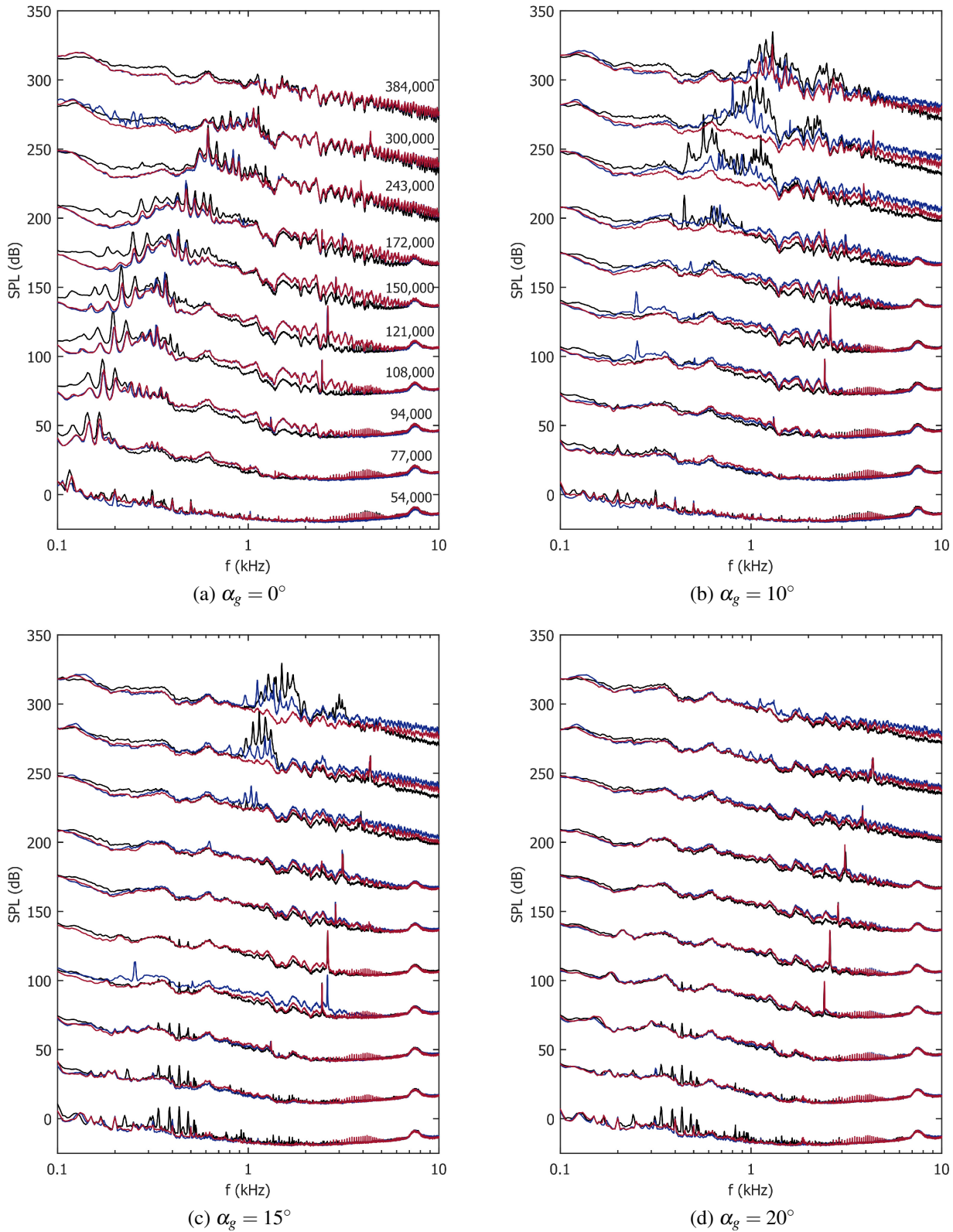
\begin{figure}[h!] 
\begin{minipage}{.48\linewidth}
\centering
\subfloat[$\alpha_g = 0^\circ$]{\label{fig: SingleMic-0deg}\includegraphics[scale=.6]{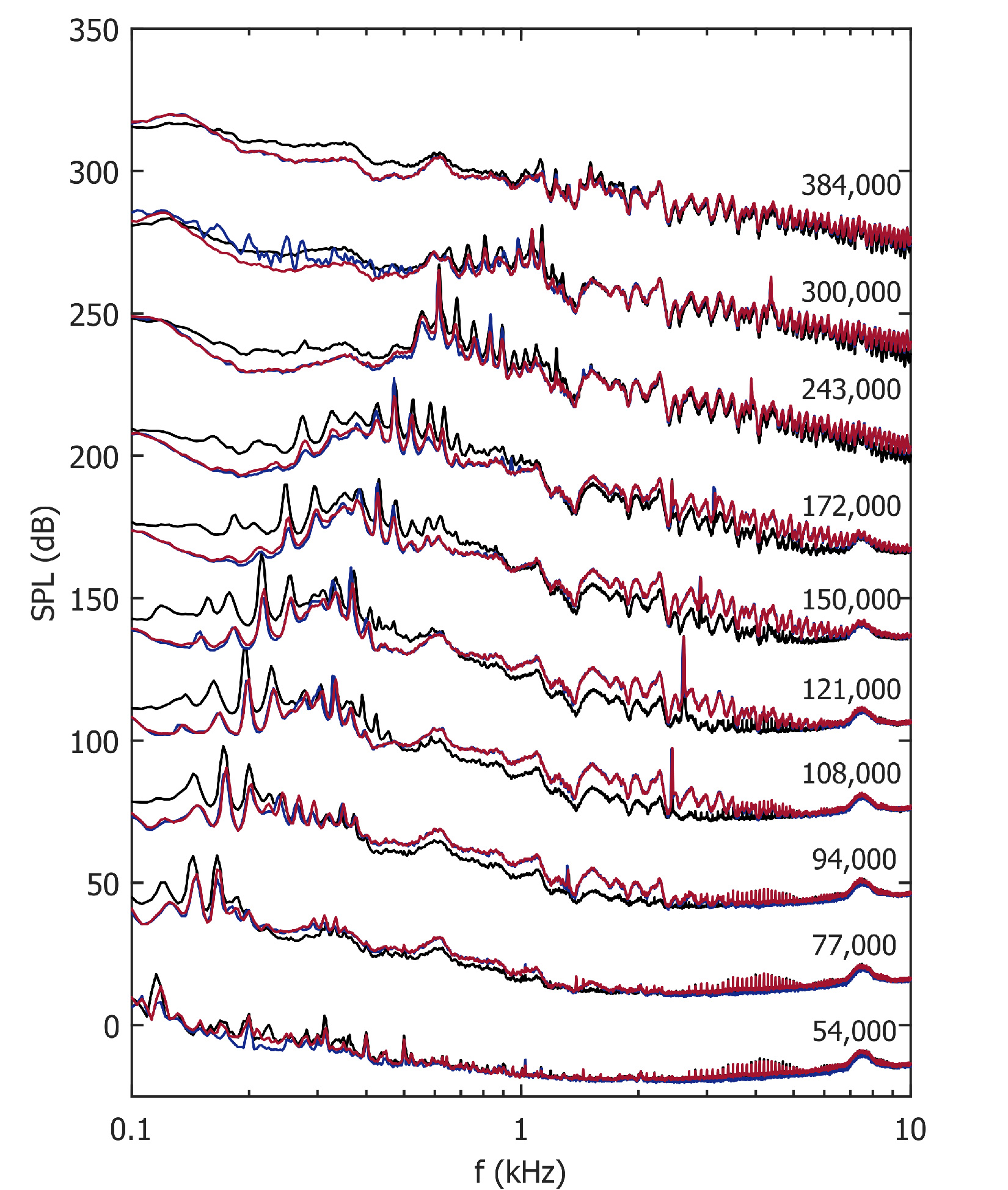}}
\end{minipage}
\begin{minipage}{.48\linewidth}
\centering
\subfloat[$\alpha_g = 10^\circ$]{\label{fig: SingleMic-10deg}\includegraphics[scale=.6]{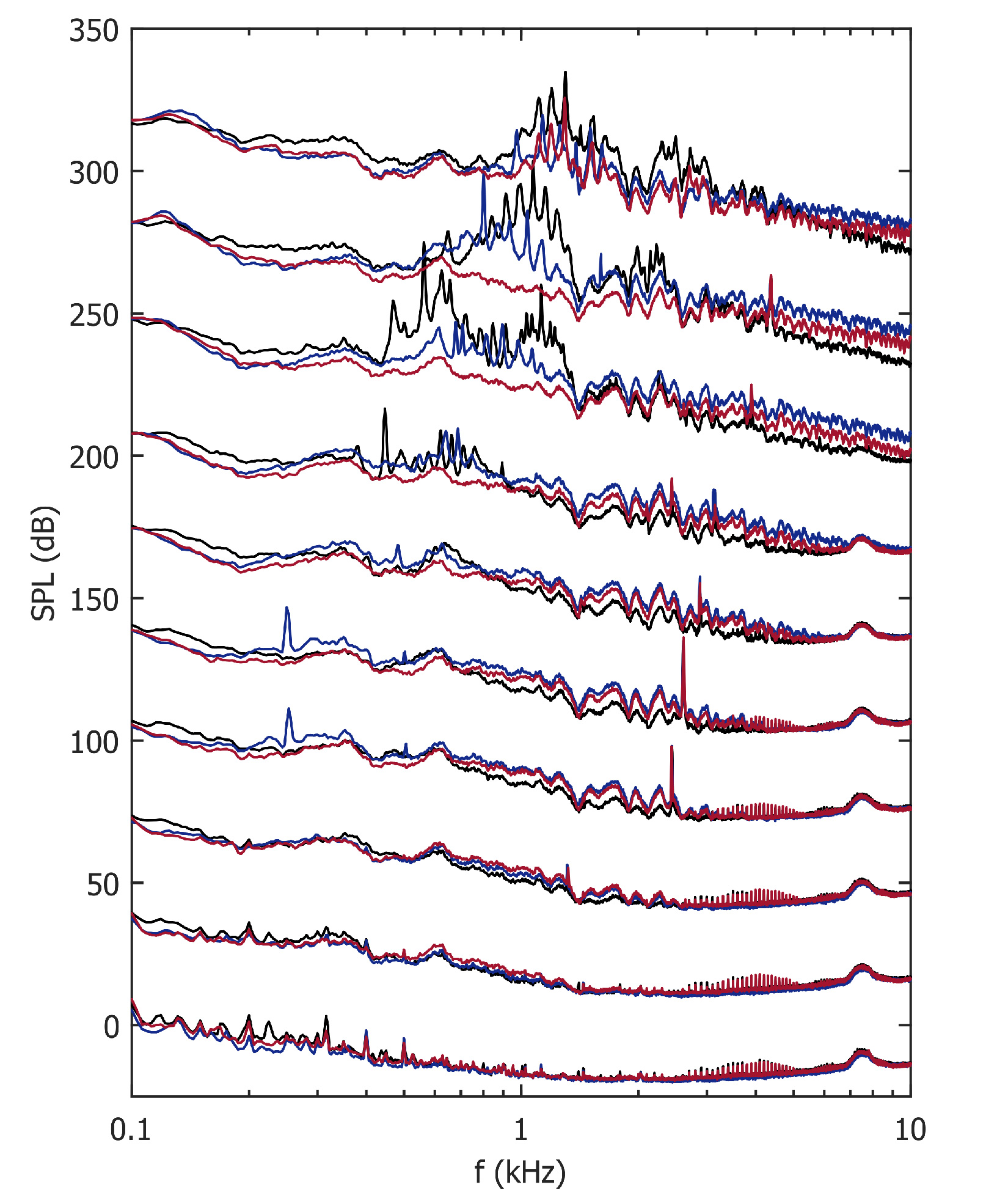}}
\end{minipage}\par\medskip
\begin{minipage}{.48\linewidth}
\centering
\subfloat[$\alpha_g = 15^\circ$]{\label{fig: SingleMic-15deg}\includegraphics[scale=.6]{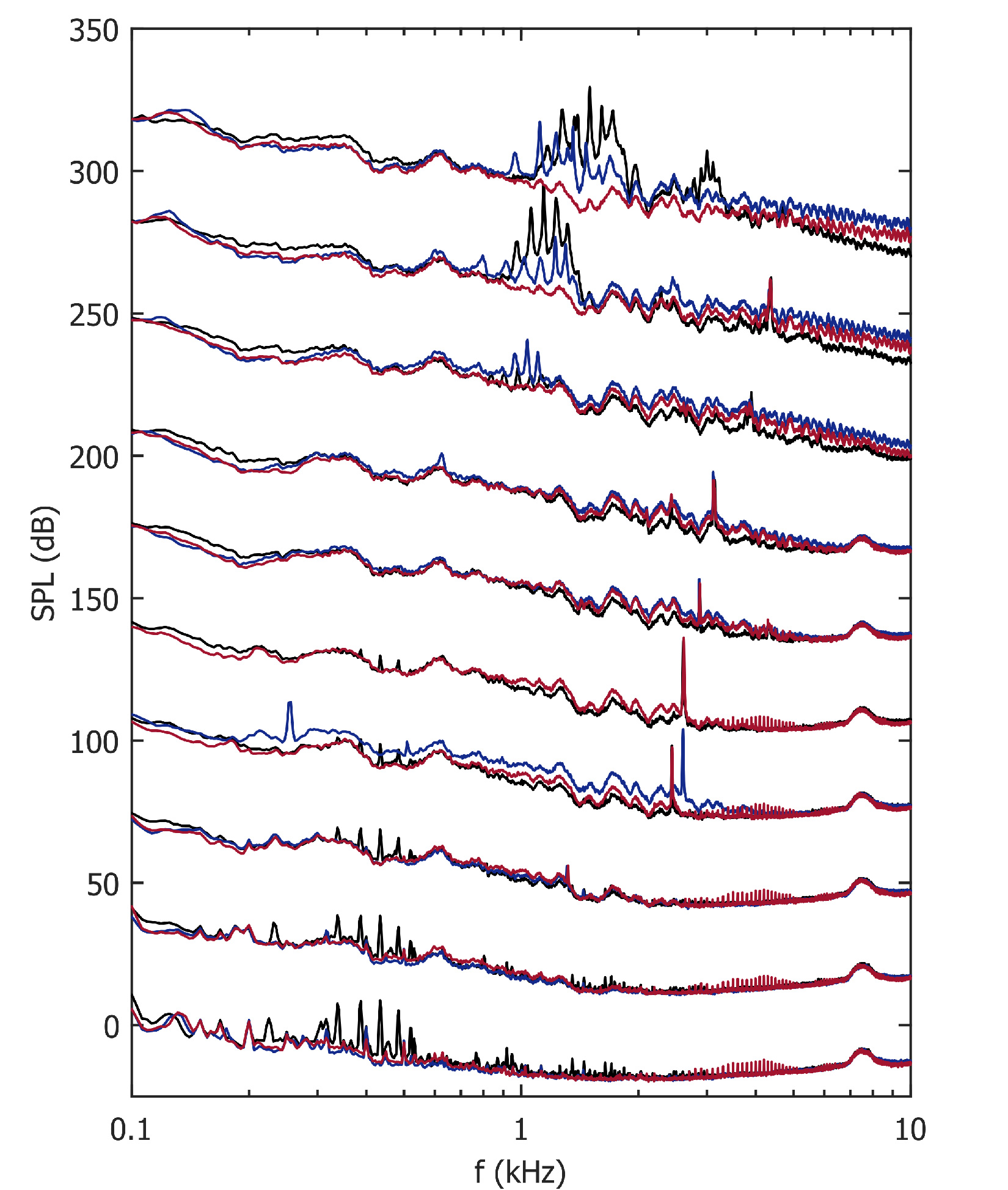}}
\end{minipage}
\begin{minipage}{.48\linewidth}
\centering
\subfloat[$\alpha_g = 20^\circ$]{\label{fig: SingleMic-20deg}\includegraphics[scale=.6]{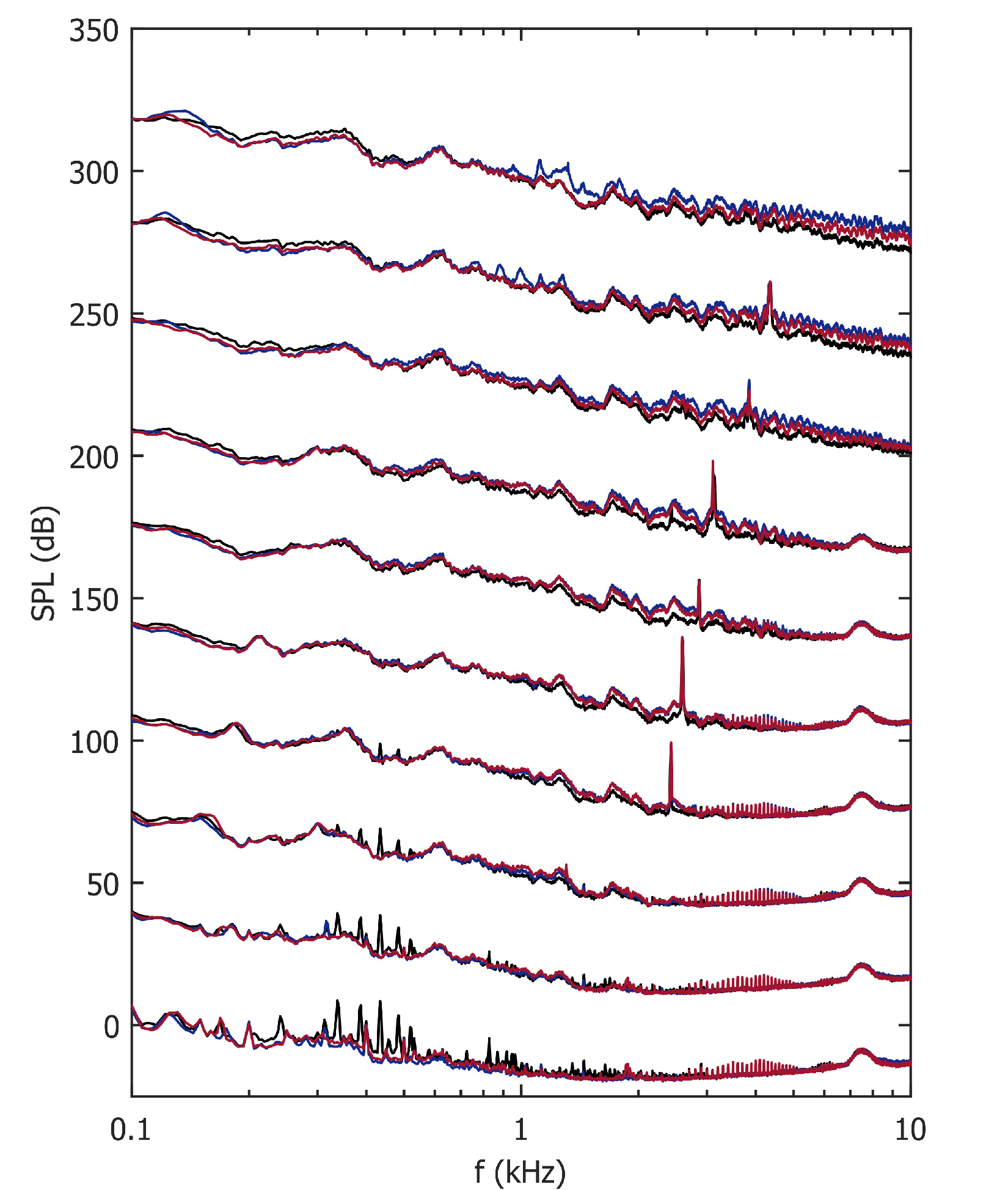}}
\end{minipage}\par
\caption{Single microphone measurements for the untripped cases. Each Reynolds number case, indicated on (a), is spaced by 30dB for clarity. Each of the angles stated are the geometric angles of attack. For each angle and Reynolds number there are three test cases: a baseline case with no flaplets (\protect\begin{tikzpicture}
\protect\tikz[baseline=-2pt] \protect\draw[solid,very thick,color=black] (0.0,0.0) -- (0.3,0.0);
\protect\end{tikzpicture}), the case where the flaplets are affixed to the pressure side (\protect\begin{tikzpicture}
\protect\tikz[baseline=-2pt] \protect\draw[solid,very thick,color=pinot] (0.0,0.0) -- (0.3,0.0);
\protect\end{tikzpicture}) and when the flaplets are affixed onto the suction side (\protect\begin{tikzpicture}
\protect\tikz[baseline=-2pt] \protect\draw[solid,very thick,color=royal] (0.0,0.0) -- (0.3,0.0);
\protect\end{tikzpicture}).}\label{fig: SingleMic}
\end{figure}

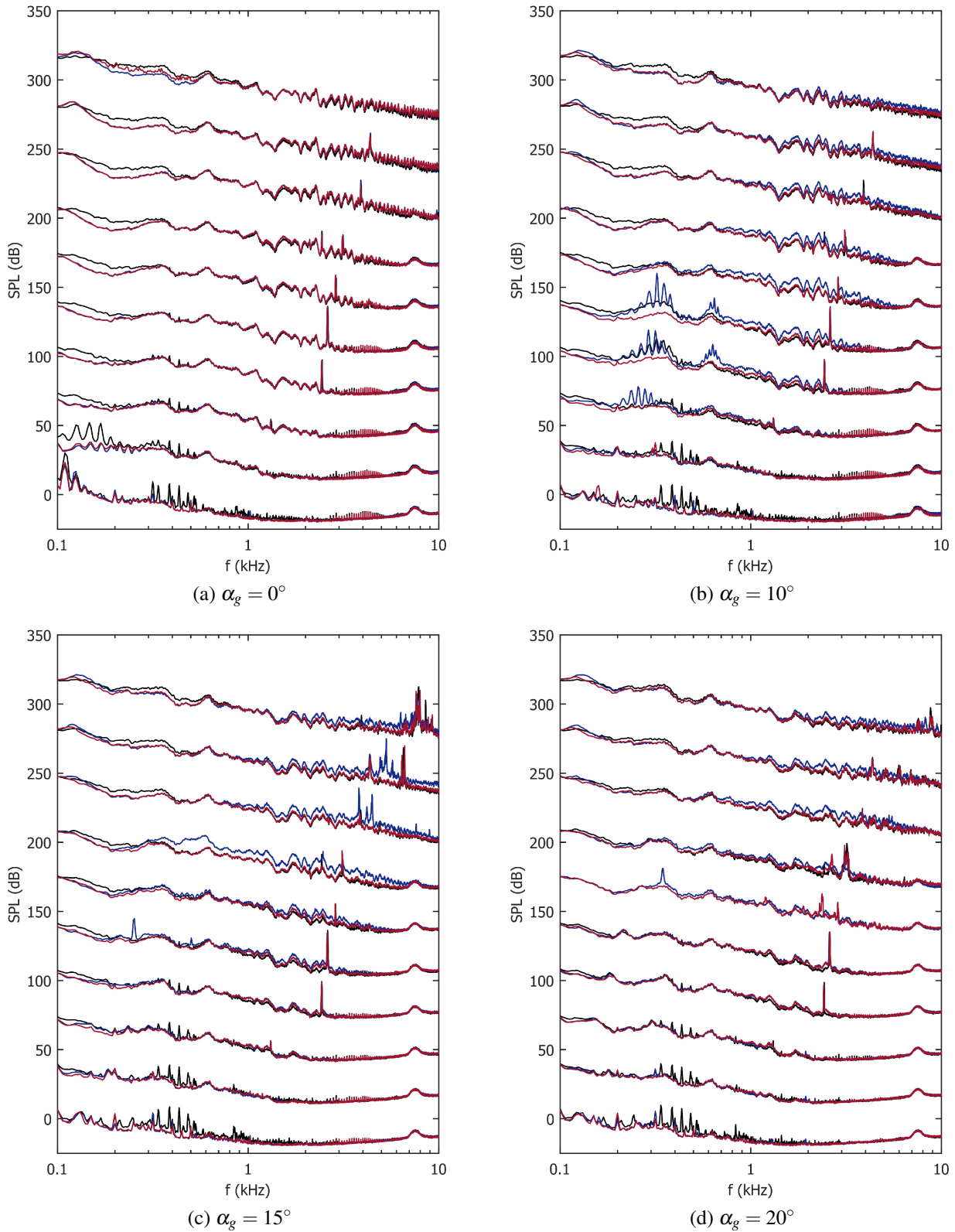
\begin{figure}[h!] 
\begin{minipage}{.48\linewidth}
\centering
\subfloat[$\alpha_g = 0^\circ$]{\label{fig: SingleMic-0degT}\includegraphics[scale=.6]{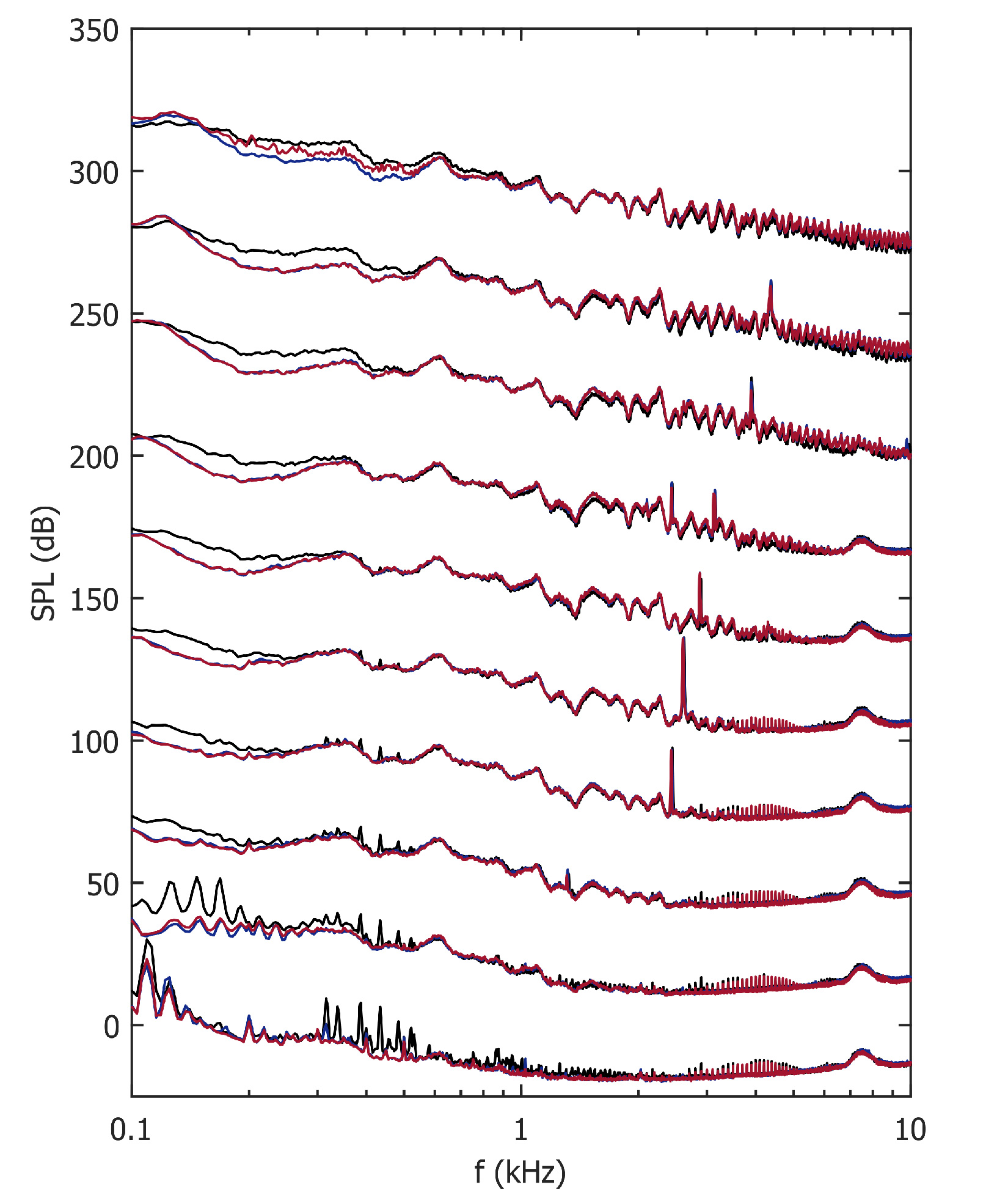}}
\end{minipage}
\begin{minipage}{.48\linewidth}
\centering
\subfloat[$\alpha_g = 10^\circ$]{\label{fig: SingleMic-10degT}\includegraphics[scale=.6]{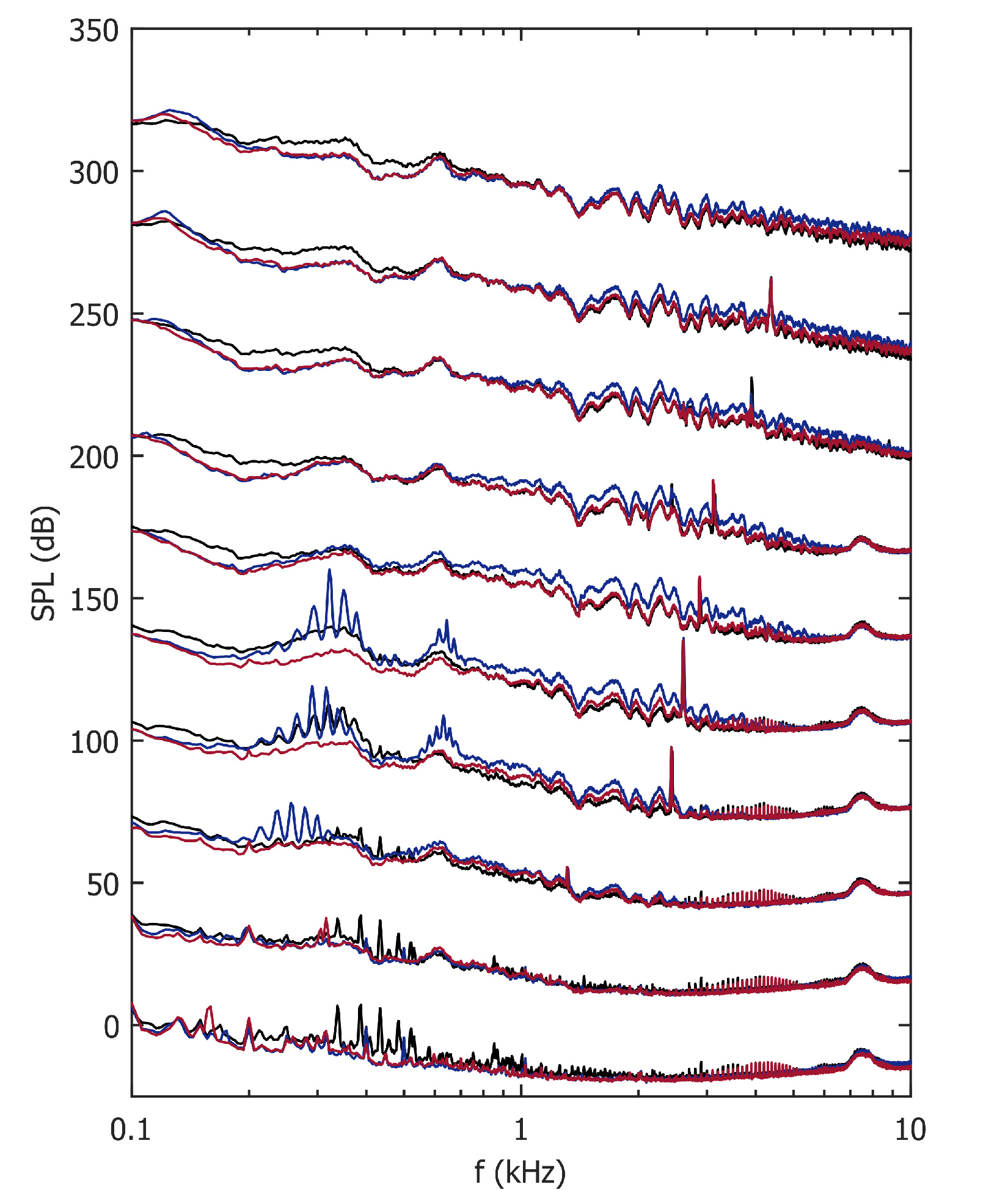}}
\end{minipage}\par\medskip
\begin{minipage}{.48\linewidth}
\centering
\subfloat[$\alpha_g = 15^\circ$]{\label{fig: SingleMic-15degT}\includegraphics[scale=.6]{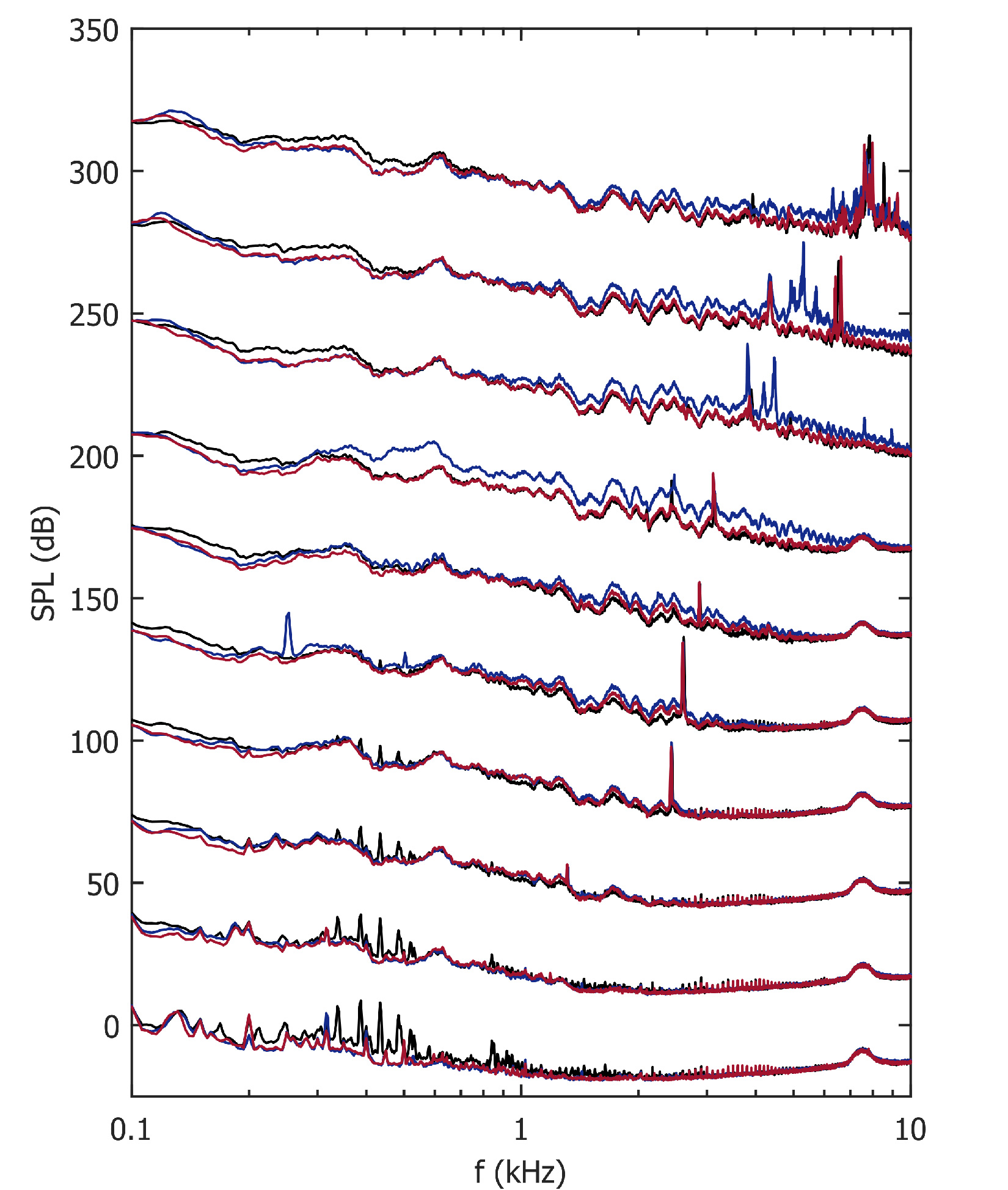}}
\end{minipage}
\begin{minipage}{.48\linewidth}
\centering
\subfloat[$\alpha_g = 20^\circ$]{\label{fig: SingleMic-20degT}\includegraphics[scale=.6]{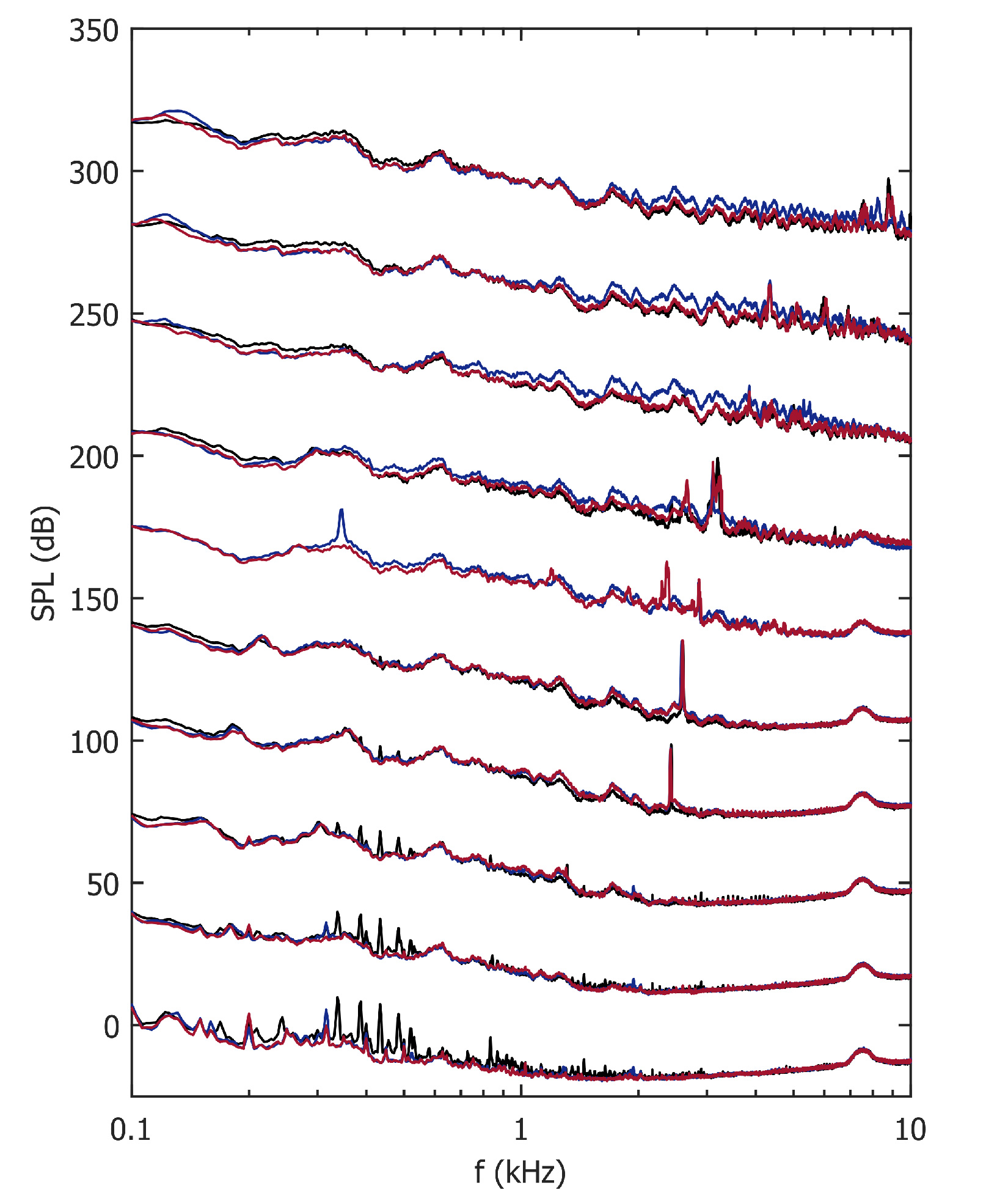}}
\end{minipage}\par
\caption{Single microphone measurements for the tripped cases. Each Reynolds number case, indicated on \ref{fig: SingleMic-0deg}, is spaced by 30~dB for clarity. Each of the angles stated are the geometric angles of attack. For each angle and Reynolds number there are three test cases: a baseline case with no flaplets (\protect\begin{tikzpicture}
\protect\tikz[baseline=-2pt] \protect\draw[solid,very thick,color=black] (0.0,0.0) -- (0.3,0.0);
\protect\end{tikzpicture}), the case where the flaplets are affixed to the pressure side (\protect\begin{tikzpicture}
\protect\tikz[baseline=-2pt] \protect\draw[solid,very thick,color=pinot] (0.0,0.0) -- (0.3,0.0);
\protect\end{tikzpicture}) and when the flaplets are affixed onto the suction side (\protect\begin{tikzpicture}
\protect\tikz[baseline=-2pt] \protect\draw[solid,very thick,color=royal] (0.0,0.0) -- (0.3,0.0);
\protect\end{tikzpicture}).}\label{fig: SingleMicTripped}
\end{figure}

Figures \ref{fig: SingleMic} and \ref{fig: SingleMicTripped} are the single microphone measurements from the microphone situated vertically above the trailing edge, see fig. \ref{fig: Exp Arr. Acc}. In fig. \ref{fig: SingleMic} the aerofoil is untripped, in order to observe the effect that the flaplets have on the LBL-TE interaction. Each Reynolds number is spaced by 30dB for clarity, their corresponding $Re_c$ is indicated on fig.\ref{fig: SingleMic-0deg}. At zero incidence, fig.\ref{fig: SingleMic-0deg}, it can be seen that for all Reynolds number a tonal peak is observed. An interesting observation can be seen in the low frequency range (0.1~kHz -- 0.4~kHz) where there is a significant reduction in the noise level across all cases once the flaplets are applied. 
There is no preference in the surface placement of the flaplets. However this is expected due to the symmetry of the aerofoil at $\alpha_g = 0^\circ$. 
The reduction that is seen is thought to be related to the vortex shedding noise which is thought that have changed due to the flaplets modifying the wake. This reduction of vortex shedding noise has been observed on a cylinder with flexible elements on the aft half of the cylinder \citep{Kamps2017a,Geyer2017}. \citet{Jodin2017} has showed that by using a similar, but active, trailing edge modification the wake structure is modified and it is this type of flow modification that is thought to be the mechanism behind the low frequency noise reduction. It can be seen that the reduction in the low frequency range has been scattered into the medium to high frequency range ($\sim$ 1~kHz). As the Reynolds number is increased, the reduction in the low frequency and the high frequency increase both reduce. This is postulated to be due to the low Eigen frequency of the current flaplet geometry and therefore are limited to a finite low frequency range for these benefits to be observed. 

\begin{figure}[h!] 
\centering
{\includegraphics[width=0.8 \linewidth]{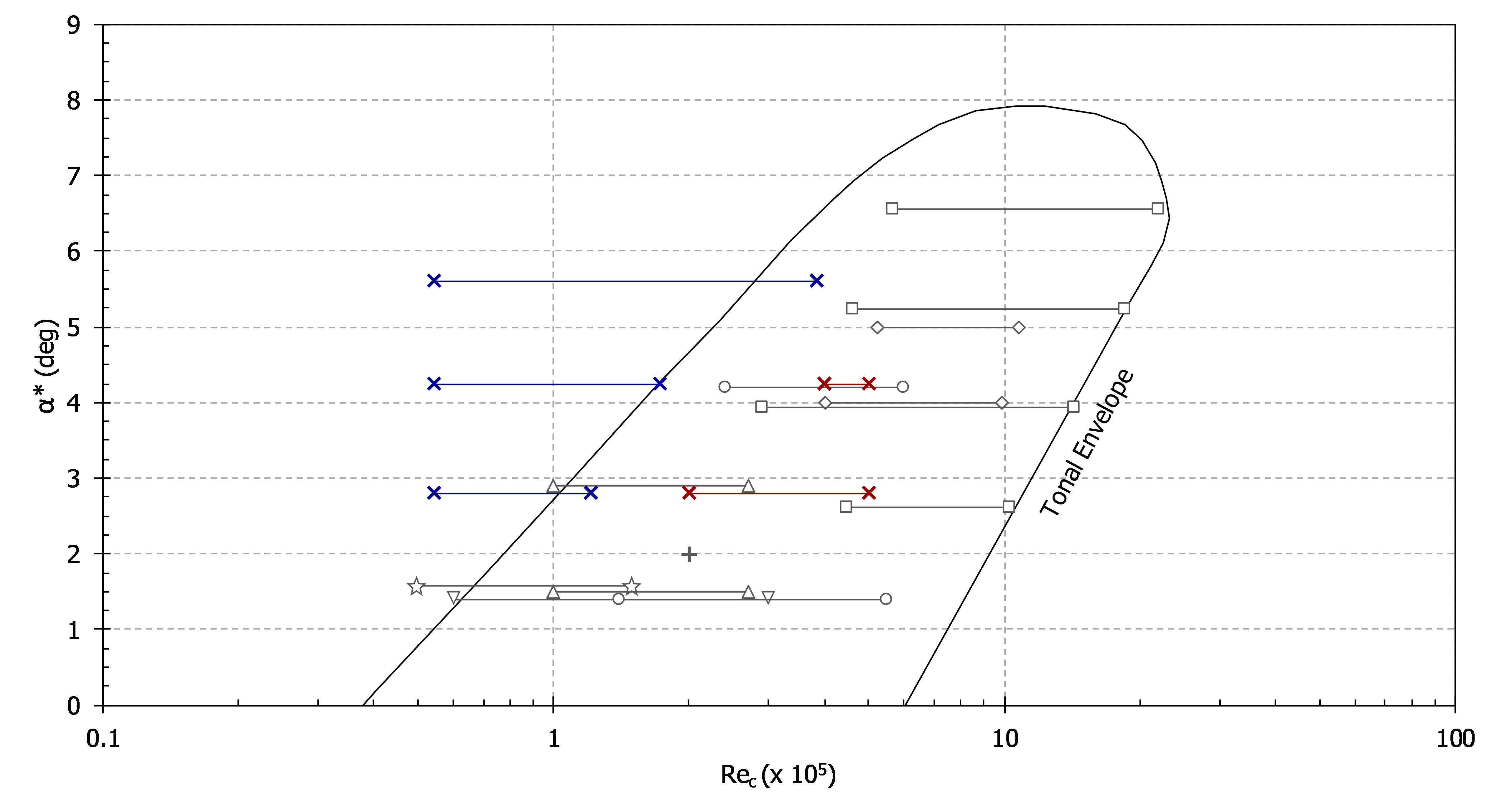}}
\caption{Comparison of previous literatures and the present study in the tonal noise envelope for the NACA 0012 aerofoil as proposed by \citet{Lowson}. The angle of attack ($\alpha^*$) is corrected using the BPM empirical correction for the open jet wind tunnel results. Neither the direct numerical simulation (DNS) or closed wind tunnel angles have been adjusted.
\protect\begin{tikzpicture}
\protect\rect{color=darkgray,fill=white};
\protect\end{tikzpicture} \citet{Paterson1972};
\protect\begin{tikzpicture}
\protect\swrect{color=darkgray,fill=white};
\protect\end{tikzpicture} \citet{Lowson};
\protect\begin{tikzpicture}
\protect\tikz[baseline=0pt] \protect\draw[solid,very thick,color=darkgray] (0.1,0.0) -- (0.1,0.2) (0.0,0.1) -- (0.2,0.1);
\protect\end{tikzpicture} \citet{Desquesnes2007};
\protect\begin{tikzpicture}
\protect\tridown{color=darkgray,fill=white};
\protect\end{tikzpicture} \citet{Inasawa2013};
\protect\begin{tikzpicture}
\protect\tikz[baseline=0pt] \protect\draw[thick,color=darkgray,fill=white] (0.1,0.1) circle (0.1);
\protect\end{tikzpicture} \citet{Chong2013};
\protect\begin{tikzpicture}
\protect\triup{color=darkgray,fill=white};
\protect\end{tikzpicture} \citet{Probsting2014};
\protect\begin{tikzpicture}
\protect\tstar{0.05}{.13}{5}{54}{semithick,color=darkgray,fill=white}; 
\protect\end{tikzpicture} \citet{Arcondoulis2018};
\protect\begin{tikzpicture}
\protect\tikz[baseline=0pt] \protect\draw[solid,thick,color=pinot] (0.0,0.0) -- (0.2,0.2) (0.0,0.2) -- (0.2,0.0);
\protect\end{tikzpicture} Present (tonal) and
\protect\begin{tikzpicture}
\protect\tikz[baseline=0pt] \protect\draw[solid,thick,color=royal] (0.0,0.0) -- (0.2,0.2) (0.0,0.2) -- (0.2,0.0);
\protect\end{tikzpicture} Present (non-tonal)}
\label{fig: Tonal Curve}
\end{figure}

As the angle increases to $\alpha_g = 10^\circ$, fig. \ref{fig: SingleMic-10deg}, a tonal peak starts to emerge at the $Re_c = 172,000$ case. As the Reynolds number increases further the tonal peaks in the baseline case increase in frequency and intensity. The tonal peaks which are observed for this angle of attack and the subsequent angles agree well with the `tonal envelope' model, fig.\ref{fig: Tonal Curve}, which was first proposed by \citet{Lowson}, when the angle is normalised using the empirical scaling factor \citep{Brooks1986}. A series of previous publications using the NACA 0012 have also been normalised, accounting for different experimental set-ups, and fall within this tonal envelope.

A particularly interesting result can be first seen in fig. \ref{fig: SingleMic-10deg}, where the placement of the flaplets on the pressure side of the aerofoil significantly reduces or removes the tonal frequencies. Whereas the placement on the suction side does not have such a profound impact, it should be noted that the peak is slightly reduced. This suppression of tonal noise is also seen in the $\alpha_g = 15^\circ$ case at the higher Reynolds number cases. At $\alpha_g = 20^\circ$, fig. \ref{fig: SingleMic-20deg}, there is no real discernible difference between the two flaplet orientations due to the test cases being outside the tonal envelope. The low frequency noise reduction can still be seen at higher incidences, however it is reduced and has a trend of reducing as Reynolds number and $\alpha_g$ increase, in a similar fashion to that of the $\alpha_g = 0^\circ$ case. 

In the case where the aerofoil is tripped, fig.\ref{fig: SingleMicTripped}, the low frequency `dip' is present in the same way as the untripped case (fig.\ref{fig: SingleMic}). This shows that this effect is more likely to be due to the wake modification and not due to boundary layer -- trailing edge interaction.    

\subsection{Linear stability analysis} \label{sec: LSA}
\begin{figure}[t!]	
\begin{minipage}{0.31\linewidth}
\centering
\subfloat[$Re_c = 243,000;~\alpha_g = 10^\circ$]{\label{fig: LSA Re 240k}\includegraphics[scale=0.6]{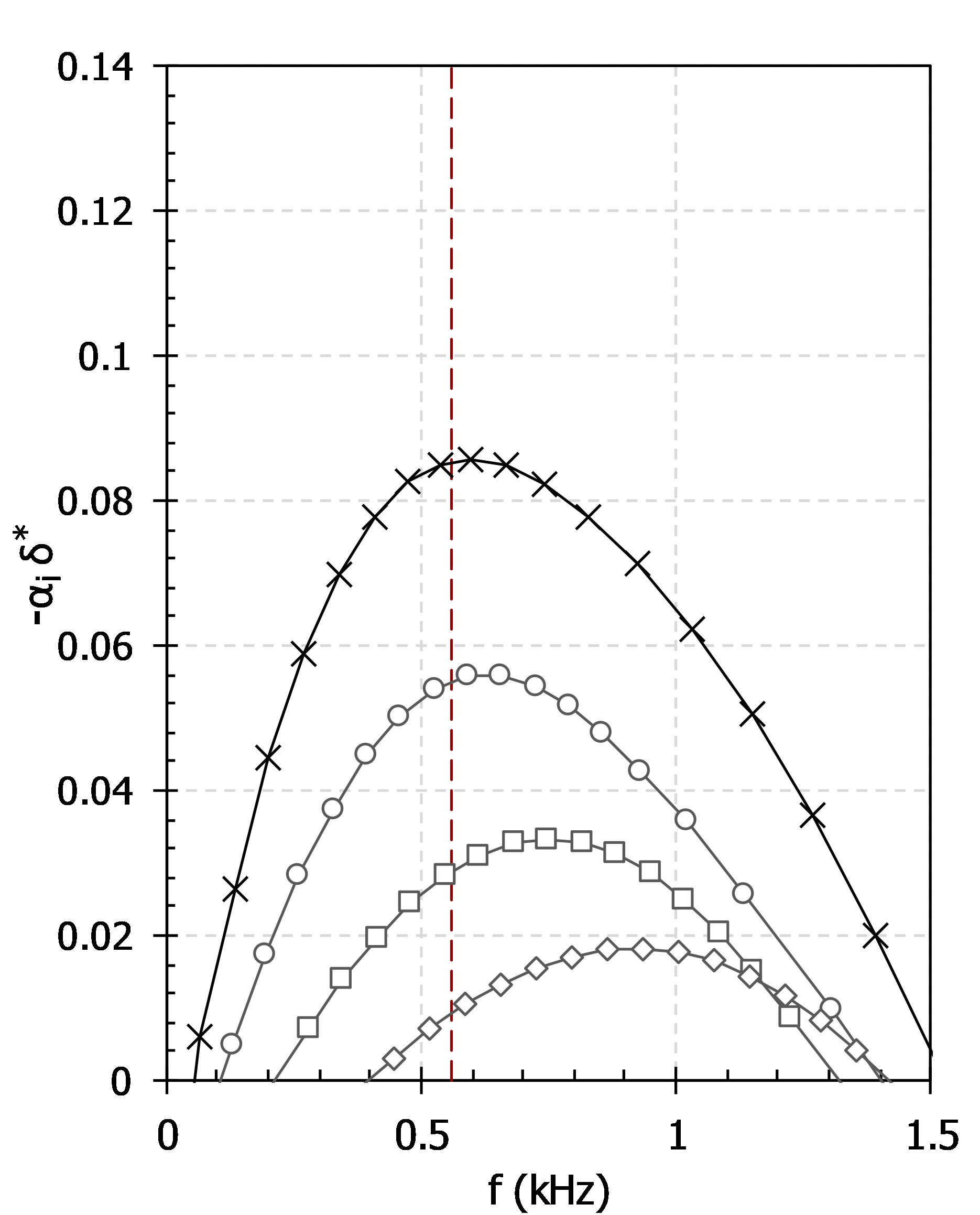}}
\end{minipage}
\begin{minipage}{0.31\linewidth}
\centering
\subfloat[$Re_c = 300,000;~\alpha_g = 10^\circ$]{\label{fig: LSA Re 300k}\includegraphics[scale=0.6]{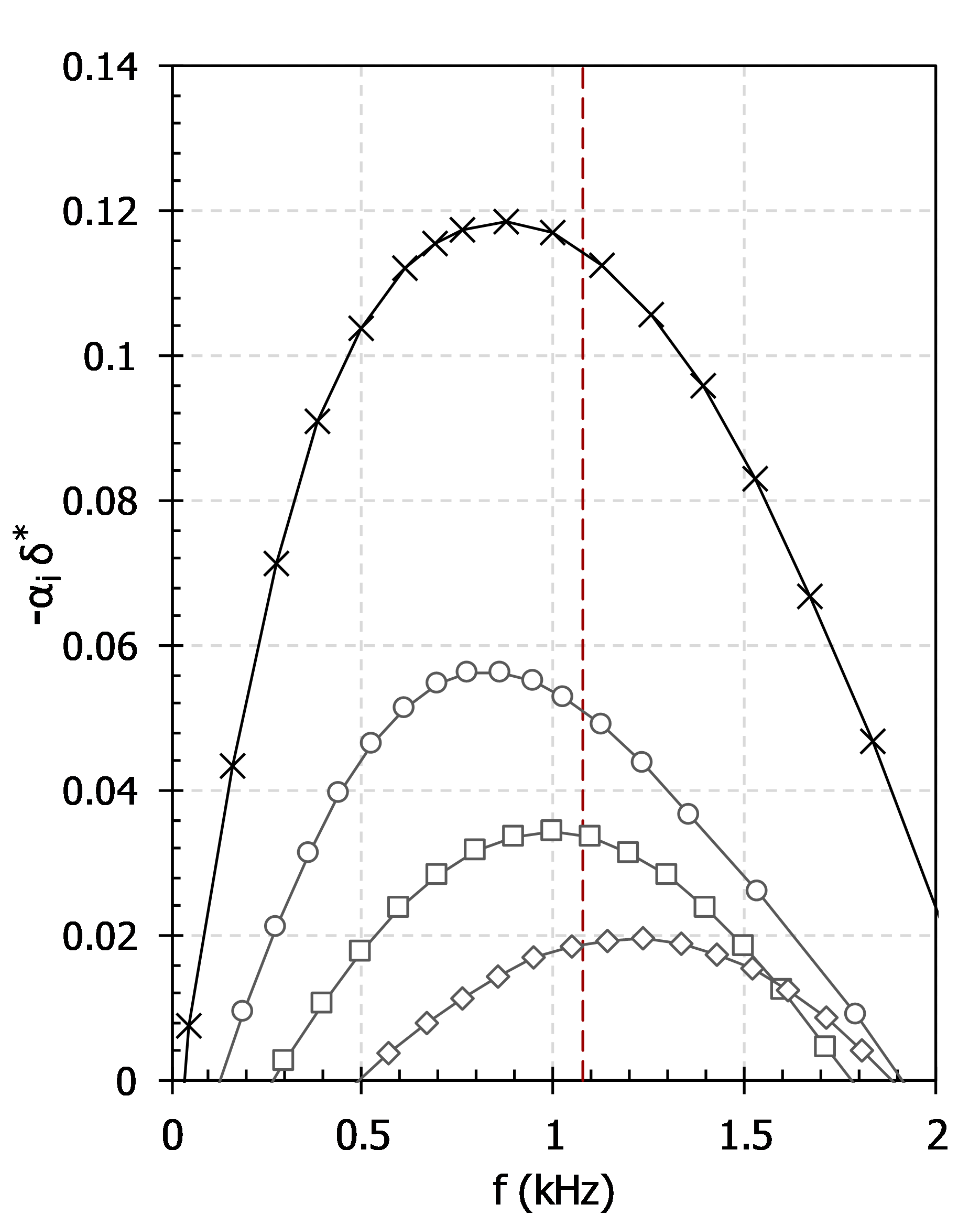}}
\end{minipage}
\begin{minipage}{0.31\linewidth}
\centering
\subfloat[$Re_c = 384,000;~\alpha_g = 10^\circ$]{\label{fig: LSA Re 385k}\includegraphics[scale=0.6]{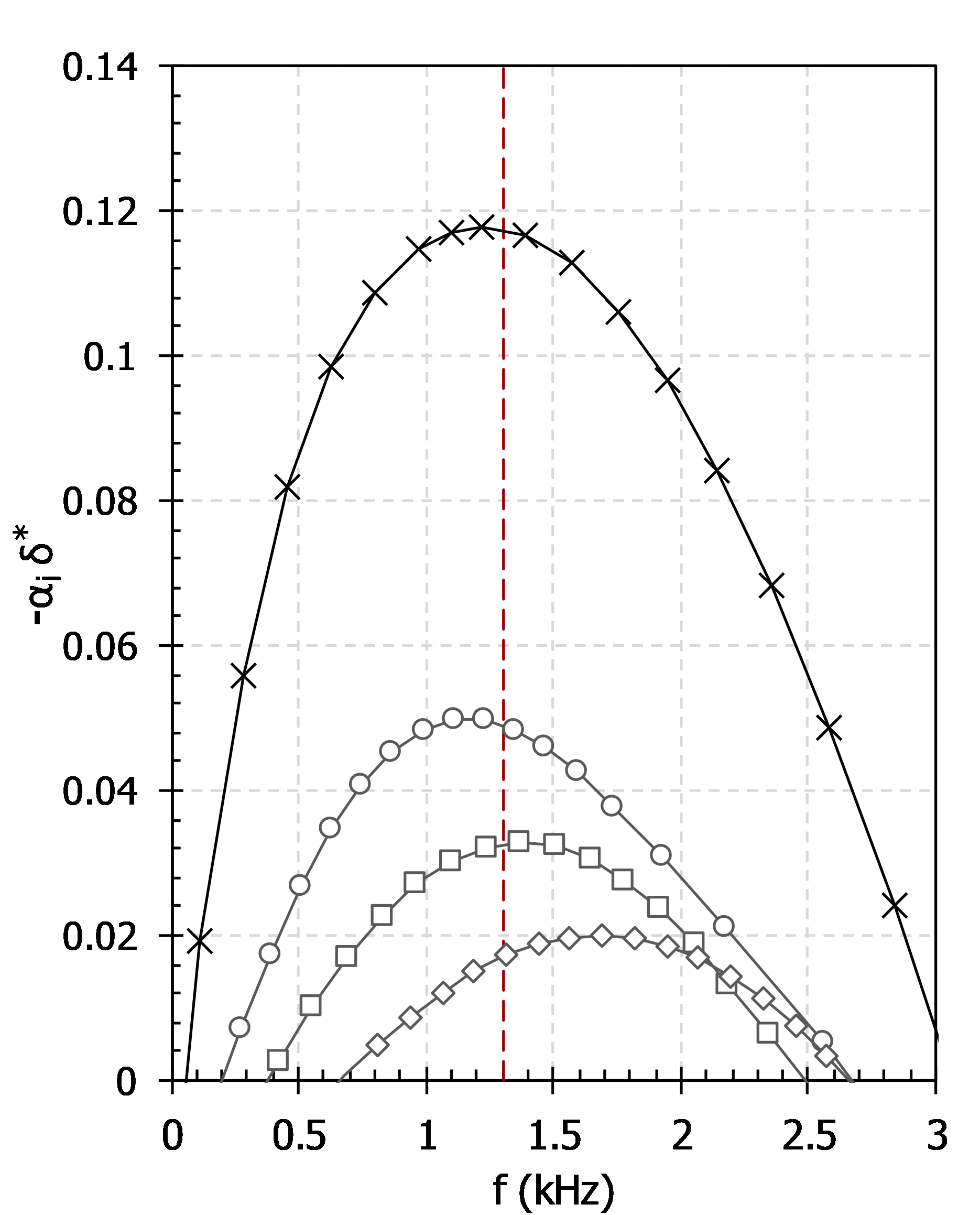}}
\end{minipage}
\caption{The spatial growth rate on the pressure side of the aerofoil at different chord wise positions (x/c) against the frequency where they occur. 
\protect\begin{tikzpicture}
\protect\swrect{color=gray,fill=white};
\protect\end{tikzpicture} 0.4c,
\protect\begin{tikzpicture}
\protect\rect{color=gray,fill=white};
\protect\end{tikzpicture} 0.5c,
\protect\begin{tikzpicture}
\protect\tikz[baseline=0pt] \protect\draw[thick,color=gray,fill=white] (0.1,0.1) circle (0.1);
\protect\end{tikzpicture} 0.6c,
\protect\begin{tikzpicture}
\protect\tikz[baseline=0pt] \protect\draw[solid,thick,color=black] (0.0,0.0) -- (0.2,0.2) (0.0,0.2) -- (0.2,0.0);
\protect\end{tikzpicture} is the position of the laminar separation bubble; (a) 0.67c, (b) 0.72c, (c) 0.76c.
\protect\begin{tikzpicture}
\protect\tikz[baseline=-2pt] \protect\draw[dashed, thick,color=pinot] (0.0,0.0) -- (0.4,0.0);
\protect\end{tikzpicture} Indicates the frequency where the tonal peak ($f_n$) is observed in the experiment, see fig.\ref{fig: SingleMic-10deg}.}
\label{fig: LSA} 
\end{figure}

As detailed by \citet{Lowson} and \citet{McAlpine1999}, the most amplified instability wave in the boundary layer prior to the separation bubble on the pressure side, is very close to that of the tonal frequency observed. 
They also stated that a separation bubble on the pressure side is a necessary requirement in the production of tonal noise on the NACA 0012 aerofoil.   
Therefore a linear stability analysis (LSA) has been carried out on the $\alpha_g = 10^\circ$ cases where the tonal peaks were observed (i.e. the top three cases on \ref{fig: SingleMic-10deg}). 
The LSA was carried out using the Airbus Callisto boundary-layer solver, a more detailed overview of the methods used in the solver can be found in \citet{Atkin2014} and the references therein.
Callisto uses a three stage process in order to obtain the pressure distribution of an aerofoil (using the Callisto Viscous Garabedian and Korn method), produce boundary profiles and then carry out a stability analysis.
In the present study the pressure distributions are created using Xfoil, which are then imported in to the QinetiQ BL2D solver.
Due to the limitations of the open jet correction factor, the true angle of attack is unknown therefore the lift coefficient was used in order to iterate the pressure distribution to obtain the correct pressure distribution for the given parameters. 
BL2D then uses a standard finite-difference, parabolic solver which is based on \citet{Horton1995} to produce boundary layer profiles up to the transition point.
These profiles are subsequently then analysed using QinetiQ CoDS which is the linear stability analysis solver.
CoDS uses an $e^\text{N}$ method to obtain the N-Factors of the boundary layer profiles, and as such it can produce an amplification curve for each boundary layer profile.
In the present results, fig. \ref{fig: LSA}, the non-dimensional spatial growth rate, $-\alpha_i \delta^*$ is plotted against modal frequency ($f$).
Where $\delta^*$ is the local displacement thickness of the boundary layer and $\alpha_i$ is the imaginary part of the spatial growth rate. 
A negative $\alpha_i$ indicates an unstable mode, hence the maxima in the $-\alpha_i \delta^*$ curves show the most unstable mode and the corresponding frequency it occurs at. 
As the current aerofoil was modelled as a semi-infinite 2D model, only Tollmein-Schlichting (T-S) waves are responsible for the instability growth.  
     
The general trend of the spatial growth rate is that as the instabilities approach the transition point (laminar separation bubble) they increase in magnitude and a more defined peak starts to emerge. At the position just before the laminar separation bubble a strong peak can be observed in the growth rate. When comparing the frequency at which this maximum growth rate occurs, to the corresponding experimental tonal peaks (see fig. \ref{fig: SingleMic-10deg}), good agreement between both sets can be seen. Therefore it can be inferred that when the flaplets are placed on the pressure side of the aerofoil, they have a strong effect on the separation bubble which is responsible for the observed tonal noise component \citep{Lowson}. By suppressing the separation bubble, and therefore the tonal component, the tonal noise feedback loop is broken.    

\subsection{Overall sound pressure level measurements}
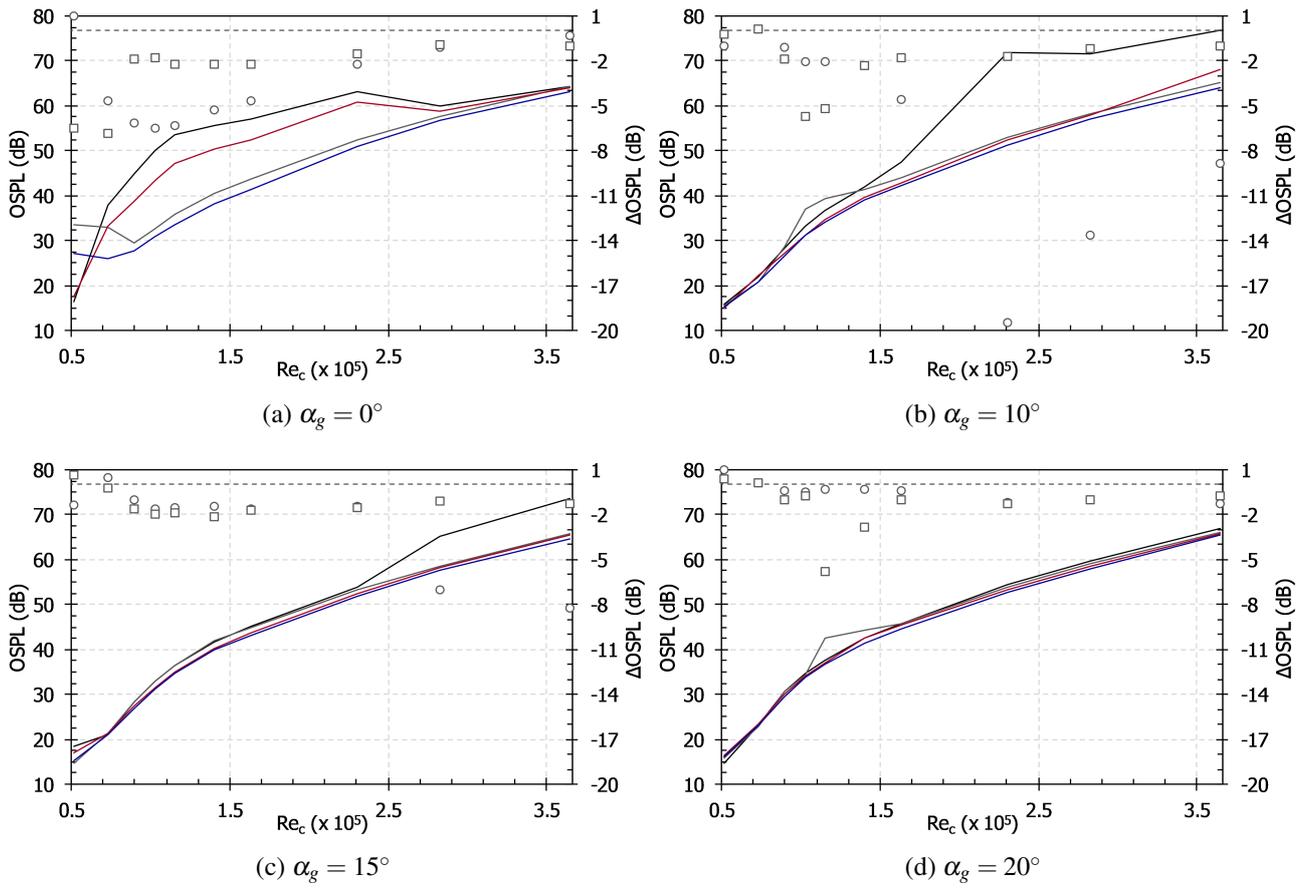
\begin{figure}[tb] 
\begin{minipage}{.48\linewidth}
\centering
\subfloat[$\alpha_g = 0^\circ$]{\label{fig: OSPL-0deg}\includegraphics[scale=.65]{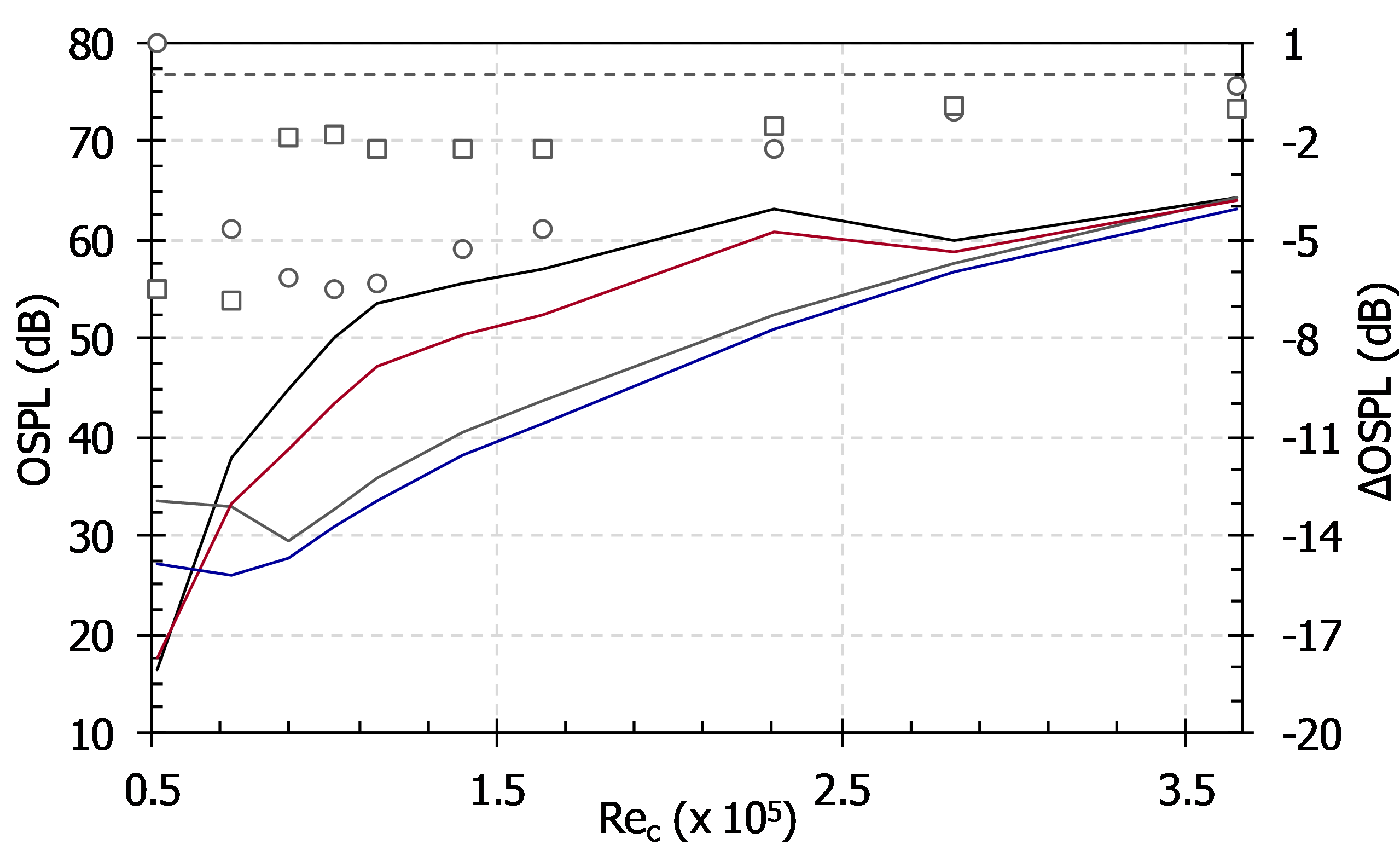}}
\end{minipage}
\begin{minipage}{.48\linewidth}
\centering
\subfloat[$\alpha_g = 10^\circ$]{\label{fig: OSPL-10deg}\includegraphics[scale=.65]{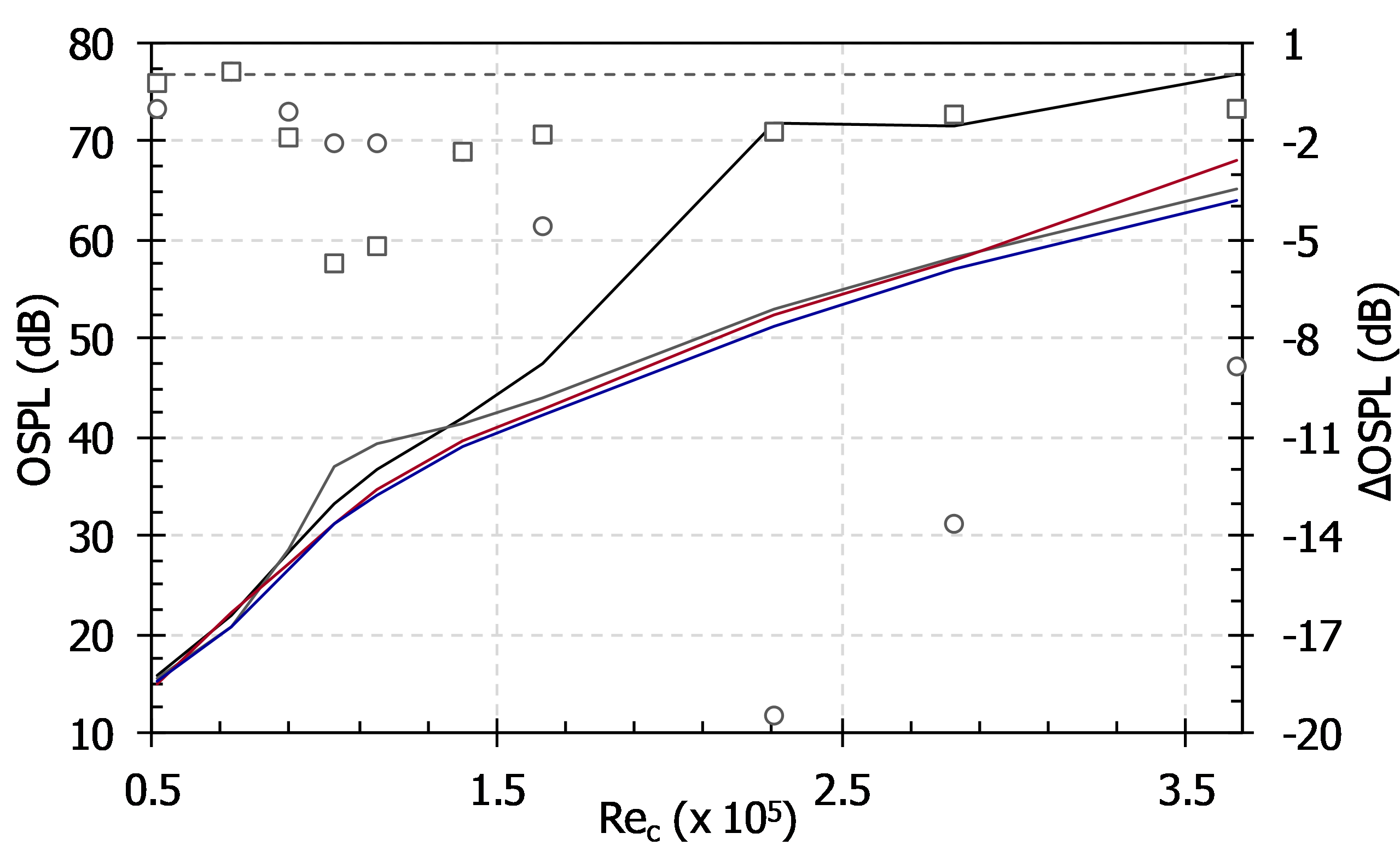}}
\end{minipage}\par\medskip
\begin{minipage}{.48\linewidth}
\centering
\subfloat[$\alpha_g = 15^\circ$]{\label{fig: OSPL-15deg}\includegraphics[scale=.65]{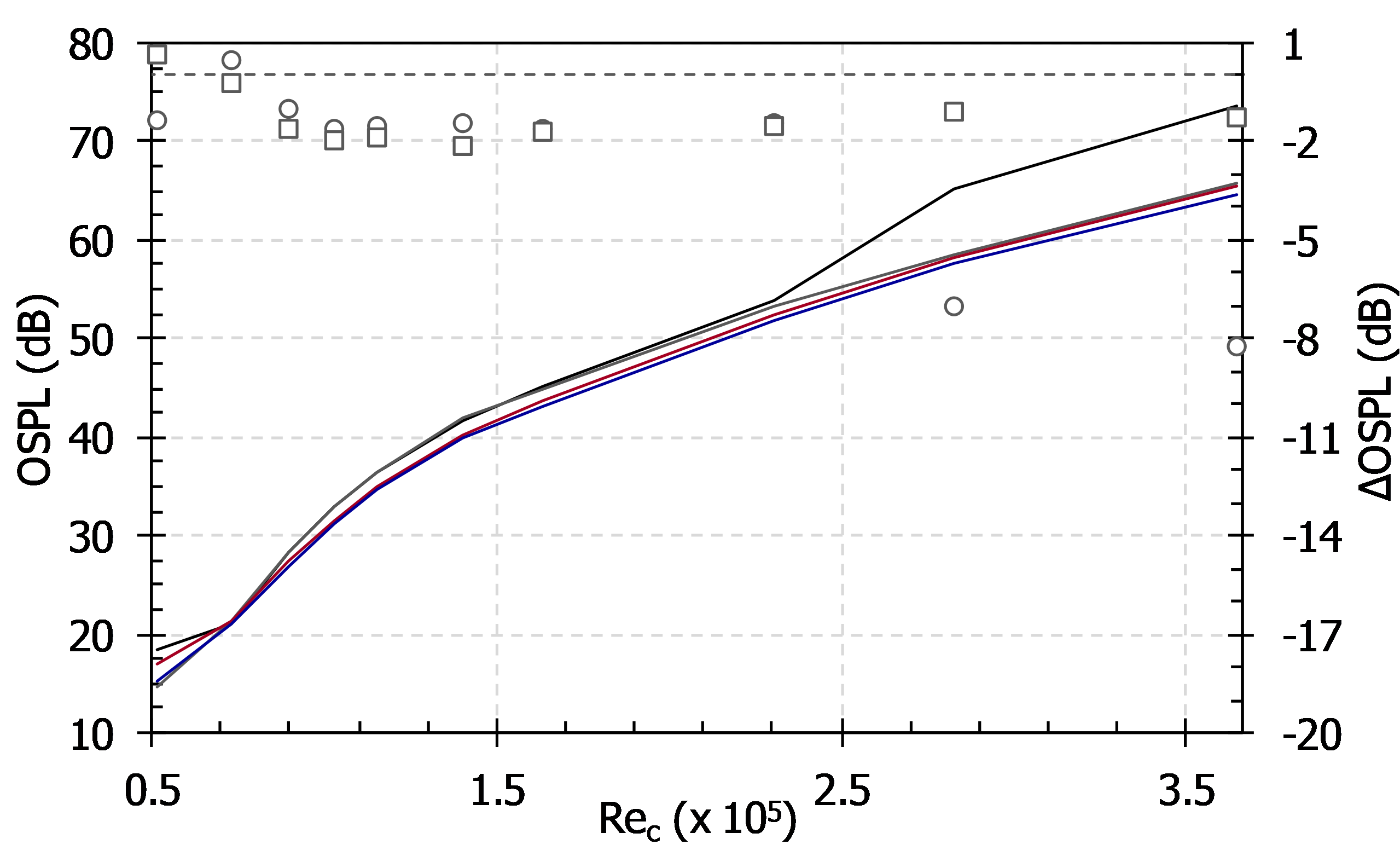}}
\end{minipage}
\begin{minipage}{.48\linewidth}
\centering
\subfloat[$\alpha_g = 20^\circ$]{\label{fig: OSPL-20deg}\includegraphics[scale=.65]{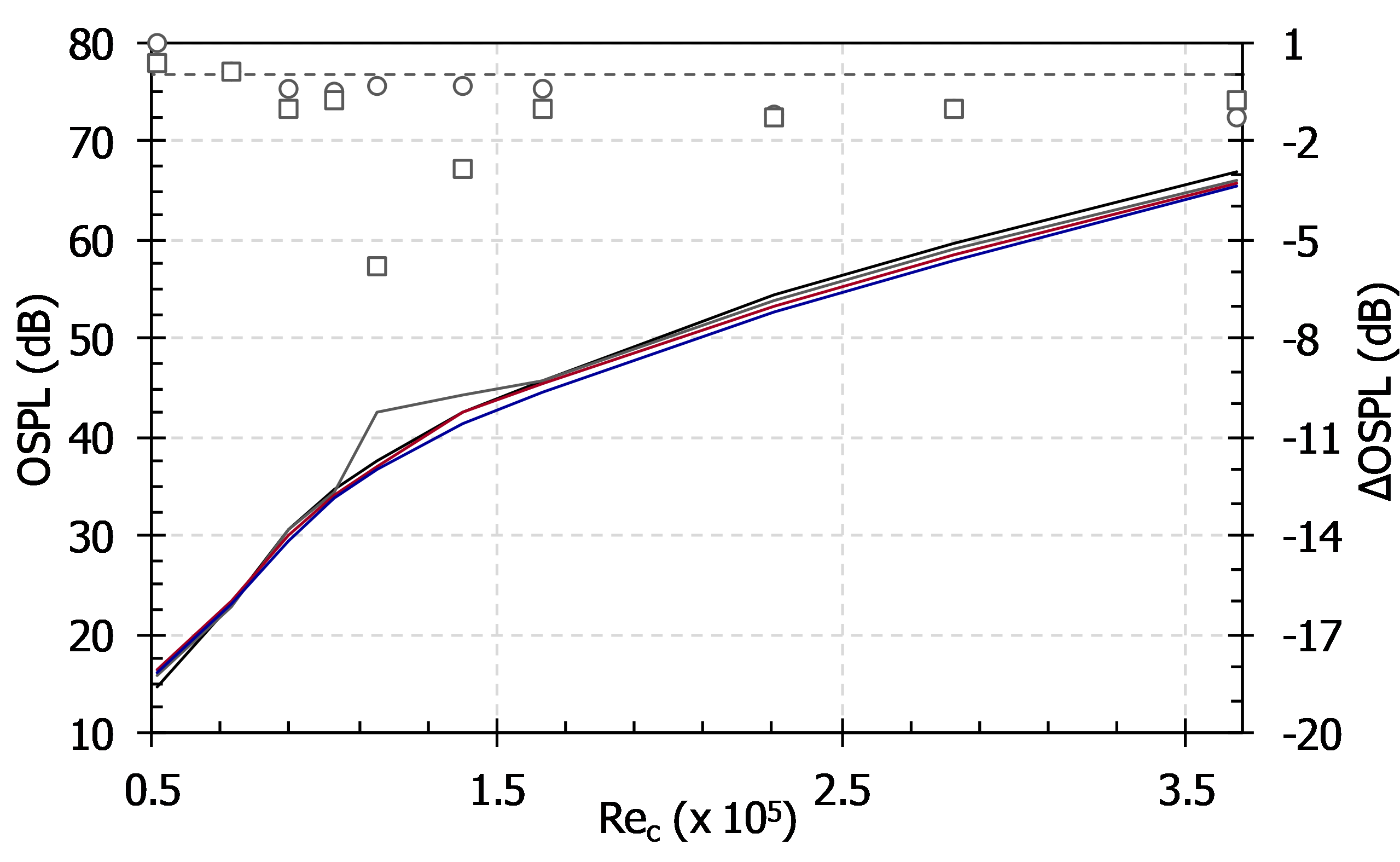}}
\end{minipage}\par
\caption{Overall sound pressure levels for the untripped and tripped baseline cases and the flaplet, pressure side mounted, cases. $\Delta$OSPL has been plotted on the second axis to yield a clear indication of the difference at each Reynolds number. Baseline untripped(\protect\begin{tikzpicture}
\protect\tikz[baseline=-2pt] \protect\draw[solid,very thick,color=black] (0.0,0.0) -- (0.3,0.0);
\protect\end{tikzpicture}), untripped flaplets pressure side mounted (\protect\begin{tikzpicture}
\protect\tikz[baseline=-2pt] \protect\draw[solid,very thick,color=pinot] (0.0,0.0) -- (0.3,0.0);
\protect\end{tikzpicture}), tripped baseline (\protect\begin{tikzpicture}
\protect\tikz[baseline=-2pt] \protect\draw[solid,very thick,color=gray] (0.0,0.0) -- (0.3,0.0);
\protect\end{tikzpicture}) and tripped flaplets pressure side mounted (\protect\begin{tikzpicture}
\protect\tikz[baseline=-2pt] \protect\draw[solid,very thick,color=royal] (0.0,0.0) -- (0.3,0.0);
\protect\end{tikzpicture}). 
\protect\begin{tikzpicture}
\protect\tikz[baseline=0pt] \protect\draw[thick,color=gray,fill=white] (0.1,0.1) circle (0.1);
\protect\end{tikzpicture} Indicates the $\Delta$OSPL of the untripped case and  
\protect\begin{tikzpicture}
\protect\rect{color=gray,fill=white};
\protect\end{tikzpicture} indicates the $\Delta$OSPL of the tripped case.
The zero line on the $\Delta$OSPL axis is shown by (\protect\begin{tikzpicture}
\protect\tikz[baseline=-2pt] \protect\draw[dashed, thick,color=gray] (0.0,0.0) -- (0.4,0.0);
\protect\end{tikzpicture}).}\label{fig: OSPL}
\end{figure}

Figure \ref{fig: OSPL}, shows the overall sound pressure level (OSPL) from the third octave bands from the whole microphone array. Here only the data for  pressure side mounted flaplets, has been presented as these have shown the most beneficial results thus far. The OSPL is a means of summing all of the acoustic tones in the signal to give one numerical value for each test case (see eqn. \ref{eqn: OSPL}). And in order to compare tripped and untripped cases, the difference between the corresponding baseline and flaplet cases has been taken (see eqn. \ref{eqn: DOSPL}). 

\begin{eqnarray}
\text{OSPL} &=& 10 \log_{10}\Bigg[\sum_{i} 10^{\text{SPL}_i/(10~\text{dB})}\Bigg] \label{eqn: OSPL}\\
\Delta\text{OSPL} &=& \text{OSPL}_\text{flaplet} - \text{OSPL}_\text{baseline} \label{eqn: DOSPL} 
\end{eqnarray}

For the $\alpha_g = 0^\circ$ case, fig. \ref{fig: OSPL-0deg}, it can be seen that in the untripped case there is a large noise reduction of $\sim$5-7~dB at the low Reynolds number range and then at higher Reynolds numbers the difference decreases. As was seen in figures \ref{fig: SingleMic-10deg} and \ref{fig: SingleMic-15deg}, there is a strong tonal component at the higher Reynolds numbers and as such can be clearly be seen as the dominating noise source in the OSPL (fig. \ref{fig: OSPL-10deg} and \ref{fig: OSPL-15deg}). Therefore the respective flaplet cases have a large $\Delta$OSPL reduction, which even extends to 20~dB at $\alpha_g = 10^\circ$. For the tripped cases, there is a much more consistent trend of noise reduction in the order of $\sim$1.5-2~dB, across all angles and Reynolds numbers, see table \ref{tab: DOSPL}. This constant reduction is due to the low frequency dip that is observed in fig. \ref{fig: SingleMicTripped}. Even though a similar effect was seen in the untripped case, fig. \ref{fig: SingleMic}, it can be seen that the increase in the high frequency range is higher, therefore the overall effect is reduced. 

\begin{table}[h!]
\centering
\begin{tabular}{c|cc|cc|cc|cc}
$\alpha_g$           		& \multicolumn{2}{c|}{0$^\circ$} & \multicolumn{2}{c|}{10$^\circ$} & \multicolumn{2}{c|}{15$^\circ$} & \multicolumn{2}{c}{20$^\circ$} \\\hline
No Trip / Trip       		& No Trip    & Trip     & No Trip     & Trip     & No Trip     & Trip     & No Trip     & Trip     \\
Average $\Delta$OSPL (dB)	& -3.62      & -2.71    & -5.49       & -2.08    & -2.48       & -1.27    & -0.41       & -1.37   
\end{tabular} 
\caption{Average $\Delta$OSPL for each of the geometric angles of attack ($\alpha_g$).} \label{tab: DOSPL}
\end{table}

\section{Conclusion}
Aeroacoustic measurements have been conducted in an open jet wind tunnel on a NACA 0012 aerofoil in order to observed the acoustic effect of a trailing edge with flexible passive flaplets (bending beam oscillators) attached in the Reynolds number range 50,000 -- 350,000 at geometric angles of attack, $\alpha_g = 0^\circ-20^\circ$. When the aerofoil is untripped strong tonal peaks are observed on the baseline case and the range that these are observed are consistent with previous literature. Once the flaplets are attached to the pressure side of the aerofoil, the tonal component is removed. This is proposed to be due to a modification to the laminar separation bubble on the pressure side, which is the key mechanism for tonal noise, and as such prevents the feedback loop described by \citet{Desquesnes2007}. If the flaplets are attached to the suction side, the tonal level is slightly reduced but still present, showing that there is a preference in flaplet-aerofoil side placement. In both the tripped and untripped case, a low frequency noise reduction and a high frequency noise increase is observed, and is seen consistently regardless of the flaplet-aerofoil side placement. The mechanism behind this low frequency reduction is thought to be due a modification of the wake, which has been seen in a previous study, on a similar trailing edge modification which uses an active oscillator \citep{Jodin2017}. This results in an overall noise reduction of $\sim$1.5 -- 2~dB over the whole range of Reynolds numbers. A further investigation using particle image velocimetry (PIV) to quantify the pressure side separation bubble modification and wake modification are currently under way to understand the mechanisms hypothesised in the present study. An additional flaplet aeroacoustic study is also currently being investigated in order to determine the effect of different flaplet geometric parameters, such as different length, width and inter-spacing between the flaplets.  

\section*{Acknowledgements}
The position of Professor Christoph Br\"{u}cker is co-funded by BAE SYSTEMS and the Royal Academy of Engineering (Research Chair no. RCSRF1617$\backslash$4$\backslash$11) and travel funding for Mr. E. Talboys was provided by The Worshipful Company of Scientific Instrument Makers (WCSIM), both of which are gratefully acknowledged. The authors would also like to thank Prof. C. Atkin, Mr. T. Backer Dirks and Mr. N. Brown for their assistance and guidance in setting up and running Callisto.

%
%

\end{document}